%% file: SOSP2026/main.tex
\documentclass[sigplan,screen]{acmart}

\copyrightyear{2026}
\acmYear{2026}
\setcopyright{cc}
\setcctype{by}
\acmConference[SOSP '26]{ACM SIGOPS 32nd Symposium on Operating Systems Principles}{September 29-October 02, 2026}{Prague, Czech Republic}
\acmBooktitle{ACM SIGOPS 32nd Symposium on Operating Systems Principles (SOSP '26), September 29-October 02, 2026, Prague, Czech Republic}
\acmDOI{10.1145/3830418.3843864}
\acmISBN{979-8-4007-2585-2/2026/09}

\makeatletter
\gdef\@copyrightpermission{
 \begin{minipage}{0.3\columnwidth}
\href{https://creativecommons.org/licenses/by/4.0/}{\includegraphics[width=0.90\textwidth]{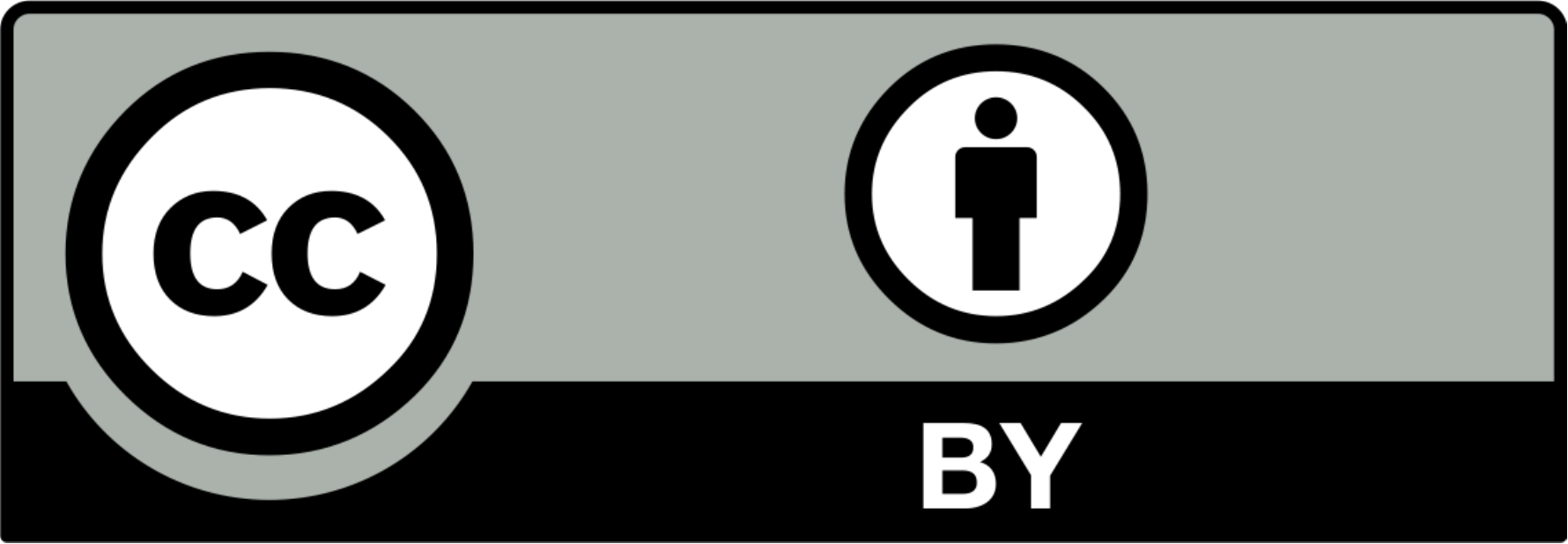}}
 \end{minipage}\hfill
 \begin{minipage}{0.7\columnwidth}
\href{https://creativecommons.org/licenses/by/4.0/}{This work is licensed under a Creative Commons Attribution International 4.0 License.}
 \end{minipage}
 \vspace{2pt}
}
\makeatother



\usepackage[]{hyperref}

\usepackage{graphicx}
\usepackage{textcomp}
\usepackage[dvipsnames]{xcolor}

\usepackage{caption}
\usepackage{cleveref}
\usepackage{soul}
\usepackage{color}
\usepackage{amsmath}
\usepackage{listings}
\usepackage{multirow,tabularx}
\newcolumntype{Y}{>{\centering\arraybackslash}X}

\usepackage{comment}
\usepackage{booktabs}
\usepackage{footnote}

\usepackage{textcomp}
\usepackage{subcaption}

\usepackage{amsthm}
\newtheorem{theorem}{Theorem} 
\newtheorem{lem}[theorem]{Lemma}    
\usepackage{algorithm}
\usepackage{algorithmicx}
\usepackage{algpseudocode}

\usepackage[inline]{enumitem}
\usepackage{tikz}
\usepackage{wrapfig}
\usepackage{float}

\newcommand*\circled[1]{\tikz[baseline=(char.base)]{
            \node[shape=circle,draw,inner sep=0.4pt] (char) {#1};}}

\newcommand{\enableComment}[1]{#1}

\renewcommand\enableComment[1]{}

\newcommand{\wz}[1]{\textcolor{blue}{ {\enableComment{[WZ: #1]}}}}

\usepackage{listings}
\definecolor{codegreen}{rgb}{0,0.6,0}
\definecolor{codegray}{rgb}{0.5,0.5,0.5}
\definecolor{codepurple}{rgb}{0.58,0,0.82}
\definecolor{backcolour}{rgb}{0.95,0.95,0.92}

\lstdefinestyle{mystyle}{
    commentstyle=\color{codegreen},
    keywordstyle=\textbf,
    numberstyle=\tiny\color{codegray},
    stringstyle=\color{codepurple},
    basicstyle=\ttfamily\footnotesize,
    breakatwhitespace=false,         
    breaklines=false,                 
    captionpos=b,                    
    keepspaces=true,                 
    numbers=left,                    
    numbersep=5pt,                  
    showspaces=false,                
    showstringspaces=false,
    showtabs=false,                  
    tabsize=2
}

\newcommand{\Cerium}[0]{Cerium}

\newcommand{\rebut}[1]{{#1}}
\newcommand{\markBlue}[1]{{#1}}

\begin{document}

\title{\Cerium{}: A Multi-GPU Framework for Terabyte-Scale Encrypted Inference}

\author{Siddharth Jayashankar}
\email{sidjay@cmu.edu}
\affiliation{%
  \department{Computer Science Department}
  \institution{Carnegie Mellon University}
  \city{Pittsburgh}
  \state{Pennsylvania}
  \country{USA}
}

\author{Joshua Kim}
\authornote{Work done at Carnegie Mellon University.}
\email{joshk@cs.utexas.edu}
\affiliation{%
  \department{Computer Science Department}
  \institution{The University of Texas at Austin}
  \city{Austin}
  \state{Texas}
  \country{USA}
}

\author{Michael B. Sullivan}
\email{misullivan@nvidia.com}
\affiliation{%
  \department{Architecture Research Group}
  \institution{NVIDIA}
  \city{Austin}
  \state{Texas}
  \country{USA}
}

\author{Wenting Zheng}
\email{wenting@cmu.edu}
\affiliation{%
  \department{Computer Science Department}
  \institution{Carnegie Mellon University}
  \city{Pittsburgh}
  \state{Pennsylvania}
  \country{USA}
}

\author{Dimitrios Skarlatos}
\email{dskarlat@cs.cmu.edu}
\affiliation{%
  \department{Computer Science Department}
  \institution{Carnegie Mellon University}
  \city{Pittsburgh}
  \state{Pennsylvania}
  \country{USA}
}

\renewcommand{\shortauthors}{ Jayashankar, et al.}

\input{Sections/Abstract}

\maketitle
\pagestyle{plain}


\input{Sections/Introduction}

\input{Sections/Background}

\input{Sections/Motivation}

\input{Sections/Design}

\input{Sections/Implementation}

\input{Sections/Evaluation}

\input{Sections/RelatedWork}

\input{Sections/Conclusion}

\input{Sections/Acknowledgements}


\bibliographystyle{ACM-Reference-Format}
\bibliography{refs}

\clearpage
\input{Sections/Appendix}

\end{document}

%% file: Sections/Abstract.tex
\begin{abstract}
Encrypted AI using fully homomorphic encryption (FHE) enables inference directly over encrypted queries, providing strong privacy guarantees. However, its computational and memory overheads have limited practical deployment. Custom FHE accelerators improve performance, but rely on advanced manufacturing technologies that limit their accessibility. GPUs offer a more widely available alternative, yet achieving ASIC-class performance on GPUs is challenging. Large models such as LLMs compound these challenges by requiring optimized kernels, terabyte-scale memory management, and efficient execution across multiple devices.

We present \Cerium{}, a multi-GPU framework for large-scale FHE inference. \Cerium{} combines a domain-specific language, an optimizing compiler, and a runtime system to automatically generate optimized GPU kernels, compress plaintext data, manage terabyte-scale memory footprints, and optimize multi-GPU communication.

Implemented on NVIDIA GPUs, \Cerium{} outperforms expert-written GPU libraries by up to 2.25$\times$ and achieves performance competitive with state-of-the-art FHE ASICs, matching the CraterLake ASIC.
It is the first GPU system to execute CKKS bootstrapping in under 10 milliseconds, achieving 7.5 milliseconds, and is the first to demonstrate encrypted inference for BERT-Base and Llama3-8B prefill in 8 seconds and 43
seconds, respectively. The \Cerium{} framework is open-source and available at \url{https://github.com/CMU-CAOS/Cerium} 
\end{abstract}

%% file: Sections/Introduction.tex
\section{Introduction}
Fully homomorphic encryption (FHE) enables computing directly on encrypted data, providing strong privacy guarantees for applications such as financial services and healthcare. 
In this paradigm, data remains encrypted throughout the computation and only the data owner can decrypt the results. FHE's confidentiality is mathematical and does not require trust in the hardware, software or any non-collusion assumptions. 
These properties make FHE attractive for secure inference, enabling users to perform AI inference over encrypted inputs on cloud-hosted models.

Unfortunately, the adoption of FHE in real-world applications has been limited by its massive performance overhead on high-end CPUs.
To address this, prior efforts~\cite{Cinnamon,Craterlake,ARK}
proposed ASICs to accelerate FHE. However, they require cutting-edge technologies (e.g., high-end nodes, CoWoS ~\cite{tsmc_cowos}, HBM) and entail high fabrication costs, making them currently impractical for widespread production and deployment. Given these constraints, datacenter GPUs emerge as a practical and accessible alternative to FHE ASICs. FHE programs are fundamentally dense vector computation and GPUs mirror several features required for FHE like massive parallelism, large
on-chip caches, and high-bandwidth memory.
However, attaining ASIC-like performance using GPUs has remained elusive.
As we will show in this paper,
this is not because the hardware is inadequate, but because attaining ASIC-like performance requires a tightly integrated software stack that reasons holistically about FHE computation --- creating optimized kernels, managing terabyte-scale memory and coordinating multi-GPU communication in ways that library-based approaches cannot achieve.

Prior works~\cite{100x,Phantom_FHE,Liberate_FHE,WarpDrive,Neo,TensorFHE,FIDESlib,cheddar} build GPU libraries for FHE that focus on kernels for low-level FHE primitives, however, they leave a substantial amount of performance untapped as FHE primitives map to very short-running kernels (5--10 us), leading to high kernel launch overheads and underutilizing hardware resources. 
State-of-the-art libraries like Cheddar~\cite{cheddar} demonstrate significantly better speedups for small FHE workloads such as ResNet-20 by manually implementing kernel fusion for a limited set of FHE primitives.
In general, achieving fast performance requires highly optimized kernels. 
However, manual optimization becomes tedious since every new application requires specialized optimizations across thousands of operations; and in practice, such difficulty leads to missed opportunities for optimization.
Large models such as BERT-Base and Llama3-8B exacerbate the problem as they require optimized FHE kernels for operators such as attention, softmax, GELU, square roots, which are far more sophisticated than the univariate polynomials supported by prior works~\cite{cheddar}.
Consequently, achieving high-performance encrypted inference at scale requires a general, automated mechanism to generate optimized GPU kernels for arbitrary FHE programs.

Beyond kernel-level optimizations, memory management emerges as a central and unaddressed challenge.
Existing GPU FHE work sidestep this issue as they focus on small models that fit entirely within a single GPU’s memory.
However, large FHE workloads like BERT-Base and Llama3-8B inference impose extreme memory pressure from three sources: (i) large encoded model weights, (ii) massive intermediate ciphertexts, and (iii) substantial evaluation-key sets. Together, they create enormous capacity requirements and heavy allocation overheads, quickly exhausting GPU and even host memory. 
Addressing this challenge fundamentally requires accounting for memory capacity, performing aggressive memory reuse, and host–device orchestration.

Finally, performance is fundamentally constrained by the limits of a single device. 
Closing the gap between FHE ASICs and GPUs ultimately requires multi-GPU scaling. 
This approach demands tightly coordinated execution, and efficient cross-GPU communication---capabilities absent from existing library-based efforts but essential for scaling FHE performance beyond the limits of a single device.

To address these challenges, this paper presents \Cerium{}---a multi-GPU framework for terabyte-scale FHE workloads. 
\Cerium{} consists of three integrated components: a domain-specific language, a compiler, and a runtime system. These components work synergistically to automatically generate optimized GPU kernels, manage memory for terabyte-scale workloads, and parallelize programs across multiple GPUs. Through this end-to-end design, \Cerium{} enables running small and large FHE workloads with a performance that matches FHE ASICs for the first time.

To automatically generate optimized kernels, our insight is that FHE primitives map to a small set of well-structured kernels. This enables defining rule-based checks on the validity of fusing kernels which can be iteratively applied to the computation graph of the entire program. To accomplish this,  \Cerium{} introduces an IR, compiler passes, and heuristics to compile FHE circuit representations to high-performance GPU kernels, thus eliminating the need to painstakingly hand-optimize application-specific kernel implementations. 

To address the memory capacity challenge, our primary insight is to take advantage of a symmetry in the algebraic structure of the FHE encoding algorithm to introduce an encoding that enables the compiler to implement $100\times$ lossless memory compression. Additionally, we design a memory layout based on the heterogeneous nature of memory accesses in FHE to enable efficient host-device memory orchestration.
 
Finally, \Cerium{} bridges the performance gap between GPUs and FHE ASICs by parallelizing FHE programs across multiple GPUs using compiler passes to reduce expensive inter-GPU communication operations, along with scheduling to overlap computation and communication.

We implement \Cerium{} on NVIDIA GPUs and evaluate it on  encrypted inference workloads of varying sizes. For small models, \Cerium{} outperforms prior expert-written hand optimized GPU libraries by $1.25\times$ on a single GPU and $2.25\times$ with multi-GPU scaling. More importantly, \Cerium{}'s performance is competitive with recent state-of-the-art FHE ASICs, matching CraterLake~\cite{Craterlake}, coming within $2.3\times$ of ARK~\cite{ARK} and $4.4\times$ of the multi-ASIC Cinnamon~\cite{Cinnamon}. \Cerium{} is the first to break the $10$ ms barrier for a CKKS bootstrapping implementation, achieving $7.5$ ms with off-the-shelf hardware. For large models, \Cerium{} reduces memory capacity requirements by $100\times$ and outperforms prior state-of-the-art GPU FHE BERT-Base inference~\cite{Thor} by $9.12\times$ and demonstrates the first FHE Llama3-8B prefill with a $43$-second TTFT.

Overall, \Cerium{} makes the following contributions: 
\begin{itemize}[leftmargin=10pt,topsep=2pt,parsep=1pt]
    \item An end-to-end multi-GPU framework for large-scale FHE machine learning inference
    \item A DSL and compiler for automatic generation of optimized GPU kernels for FHE programs
    \item A plaintext encoding that reduces memory capacity requirements for large models by over $100\times$ by exploiting algebraic structure
    \item Compiler optimizations to reduce inter-GPU traffic by $44\%$
    \item A runtime system for kernel scheduling and memory management for efficient multi-GPU FHE orchestration
    \item An evaluation of the framework on small and large FHE workloads demonstrating that current GPUs can get within $4.4\times$ the performance of state-of-the-art FHE ASICs
    \item The first to perform BERT-Base inference in 8.8 seconds and Llama3-8B prefill in 43 seconds, respectively
    \item Open Source: The \Cerium{} framework is available at \\ \url{https://github.com/CMU-CAOS/Cerium} 

\end{itemize}

%% file: Sections/Background.tex
\section{Background}
In this section, we provide background on the threat model and FHE computation for encrypted AI inference using CKKS ~\cite{ckks}, a popular FHE scheme for computing on real-valued data. 

\noindent
\subsection{Threat Model} \markBlue{
We assume a threat model similar to that of prior two-party encrypted inference work~\cite{CryptoNets,ResnetFHE,Thor,Nexus}. 
In this model, there are two parties: a user and a model service provider.
We assume that the two parties are honest-but-curious.
A user encrypts a query and sends the resulting ciphertexts to a server for inference. The server holds the model weights in plaintext and evaluates the model over the encrypted query. The user's query, and all values derived from it, receives cryptographic confidentiality without requiring trust in server-side hardware or software.
We assume that the model architecture is public. Consequently, information exposed by the execution pattern, including the input size, number of generated tokens, identity or architecture of the queried model, and other metadata inferable from the size or structure of the computation, is outside the scope of our threat model. Furthermore, as the client receives outputs of the model inference, attacks such as model stealing are out of scope.
Finally, FHE provides confidentiality but does not, by itself, guarantee the integrity or verifiability of the computation. Protecting against a malicious server that returns incorrect results is therefore also outside the scope.
}

\subsection{CKKS Background}

\begin{figure}[ht]
\centering
\includegraphics[width=0.95\linewidth]{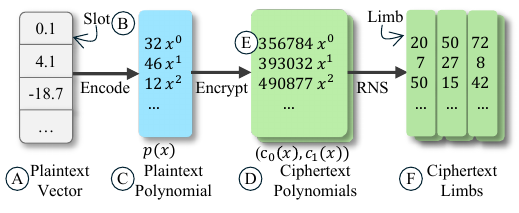}
\vspace{-2mm}
\caption{Encoding and Encrypting Values in RNS-CKKS}
\Description{}
\vspace{-2mm}
\label{fig:Background}
\end{figure}

\Cref{fig:Background} depicts the encryption steps in CKKS~\cite{ckks}, an FHE scheme for encrypted computing on real values.

\noindent
\textbf{Slots, Encoding and Encryption.}
CKKS operates over modular polynomial rings, enabling batching vectors of real values \circled{A} into a single plaintext or ciphertext. Each entry in this vector is referred to as a slot \circled{B}. Encoding converts this vector into a plaintext polynomial \circled{C} and encryption converts it to a ciphertext \circled{D}, which is a pair of modular polynomials.

\noindent
\textbf{Homomorphic Operations.}
CKKS supports three primitive homomorphic operations: (i) addition, (ii) multiplication, and (iii) cyclic rotation, that operate slotwise on the encrypted plaintext vector. All higher-level computation is expressed as circuits composed of these primitives. This restrictive model has two major implications for ML workloads.
First, CKKS cannot natively express any nonlinear function. Therefore nonlinearities like division, ReLU, max, softmax, SiLU, etc., require the use of polynomial approximations whose degree and precision must be tailored to model accuracy requirements.
Furthermore, CKKS lacks a direct indexing operator for accessing individual slots and thus cannot natively support tensor operations. Instead, tensor values have to be ``packed'' into slots and operations implemented using combinations of cyclic rotations, multiplications, and additions. This makes circuit implementations of tensor operations sensitive to factors like layer dimensions, convolution strides, batch sizes, token counts etc. As a result, FHE circuits for ML models lack generality and are tightly coupled to specific parameters making it hard to hand craft optimized implementations for all these cases.

\noindent
\textbf{RNS and Limbs.}
Polynomial coefficients \circled{E} are very large integers, often thousands of bits. Performing modular arithmetic directly on such values is inefficient. In practice, the ciphertext modulus is chosen as the product of a set of machine word-sized primes and the Residue Number System (RNS)~\cite{RNS} is used to decompose modular arithmetic over this large modulus into modular arithmetic over its prime factors, which form the RNS basis. Limbs \circled{F} refer to the modular residues of a polynomial over the RNS basis. Operations over limbs are the building blocks of practical CKKS implementations, and as most limb operations are data parallel in nature, RNS-CKKS becomes suitable for GPU acceleration.

\noindent
\textbf{NTT.} The Number Theoretic Transform (NTT) is the analog of the Fast Fourier Transform (FFT) in a prime number field. Similar to FFT, NTT performs a convolution to speed up modular limb polynomial multiplication by transforming it from the
coefficient domain to the evaluation domain. The inverse NTT (INTT) reverses this transformation. By default, all limbs can be assumed to be in the evaluation domain.

\noindent
\textbf{Bootstrapping.}
Every ciphertext has an intrinsic finite multiplicative budget that limits operations. Bootstrapping is an expensive ciphertext maintenance operation that refreshes the multiplicative budget of a ciphertext, enabling unbounded computation.  
Since bootstrapping is fundamental to all FHE computation, it is vital that it runs as fast as possible. 

\noindent
\textbf{Evaluation Keys.} Evaluation keys (evalkeys) are auxiliary data required by the keyswitching ciphertext maintenance operation. The total size of all evaluation keys in ML workloads is typically on the order of 10--50 GB. 

%% file: Sections/Motivation.tex
\section{Motivation}
In this section, we motivate the need for a holistic compiler and runtime-based approach for FHE programming on GPUs by outlining four critical, intertwined challenges that manual or library-based methods cannot adequately solve.

\subsection{Optimized Kernels are Application Specific}

FHE programs are expressed as circuits of homomorphic operations over ciphertexts. While this abstraction is meaningful for reasoning about the program, it is not the appropriate granularity for an optimized GPU implementation as FHE primitives map to short-running kernels (5-10~us) and thus severely under-utilize the GPU due to limited parallelism for latency hiding, excessive inter-kernel data communication and, kernel launching overheads.

Achieving high performance requires combining multiple operations into larger ``fused'' kernels to reduce memory transfers and better utilize GPU hardware~\cite{cheddar}.
However, this process is extremely tedious and requires deep expertise in both GPU and FHE programming.
Fundamentally, the fusion implementation is a function of the computation graph of kernels and extensive fusion 
results in the fused kernels becoming application-specific, and thus has to be repeated for every new application.

Cheddar~\cite{cheddar} manually developed a limited set of fused kernel for patterns that are mostly tailored to small workloads. For instance, Cheddar has limited support for fusion within a univariate polynomial and lacks support for fusion across multiple polynomials. 
Large workloads like BERT-Base and Llama3-8B contain operations like ciphertext - ciphertext Matmul, softmax, square root, GELU that are far more complex and not supported by Cheddar. While the complexity opens the door to more fusion opportunities; it also makes manual fusion difficult to perform. 

Resolving this problem requires a compiler that can automatically generate optimized kernels from FHE circuit specifications, performing domain-specific optimizations and fusion while also enabling generality across applications.

\subsection{Encrypted AI Dramatically Stresses Memory Capacity and Management}
\label{sec:mm_motivation}

FHE dramatically increases memory requirements, making memory management a first-order concern. Repeated runtime allocation and deallocation can introduce overheads of up to $100\%$, degrading performance even for small workloads. Memory capacity is an additional challenge for encrypted AI because model weights must be packed into plaintext polynomials and represented in RNS form, resulting in a multiplicative increase in their memory footprint.

\begin{figure}[ht]
\centering
\includegraphics[width=0.88\linewidth]{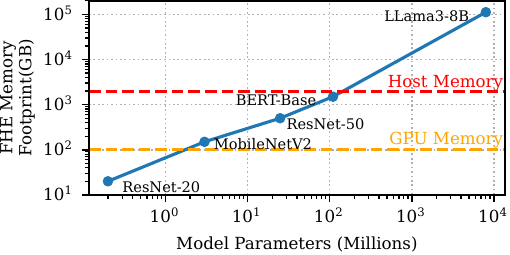}
\vspace{-3mm}
\caption{FHE Memory Footprint vs Model Parameters}
\Description{}
\label{fig:ModelMemoryCapacity}
\vspace{-4mm}
\end{figure}

~\Cref{fig:ModelMemoryCapacity} shows the memory capacity requirements of encrypted inference for models of varying sizes. Even smaller models like ResNet-50 and MobileNet have memory footprints exceeding several hundreds of gigabytes~\cite{FHEMemory}, much larger than what can fit in a GPU's memory. Encrypted LLM inference like BERT-Base or Llama3-8B requires terabytes of memory and becomes far too large to fit in GPU memories and even host main memories. BERT-Base, for instance, needs 1.5~TB of memory, and Llama3-8B requires 112~TB. Beyond inputs, encrypted LLMs further contain large amounts of intermediate ciphertext values that further stress the memory capacity on the GPU.
Previous GPU-FHE works~\cite{WarpDrive,100x,cheddar,Neo} focus on small models such as ResNet-20 and therefore avoid memory capacity challenges, or estimate end-to-end inference runtime by adding the runtime of each component~\cite{Nexus,Thor}. 

Efficient execution, therefore, requires memory management and orchestration of data movement between the host and GPUs.
These requirements are well-suited to a compiler and runtime system that can perform memory management tasks such as liveness analysis, reuse memory buffers, and orchestrate data movement between host and GPUs.

\subsection{Kernel Launch Overheads are Significant}
FHE programs, even when driven by aggressively hand-optimized kernels, still suffer from substantial performance overheads rooted in kernel launch latency from online host to device interaction.
Addressing these overheads requires asynchronous programming techniques that specify the computation as a graph of kernels that can be launched on the GPU in a single operation. 
However, manually creating such graphs, e.g., through CudaGraphs, is difficult because it requires precise tracking of kernel boundaries, memory lifetimes, and aliasing to avoid data races.
As a result, this becomes a dynamic and cross-cutting task that is best handled by a compiler that can analyze dependencies among kernels to generate correct, modular, and reusable execution graphs.

\begin{figure*}[ht]
\centering
\includegraphics[width=0.98\linewidth]{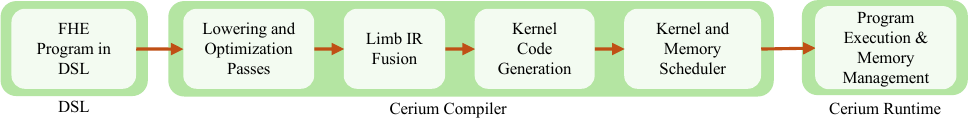}
\vspace{-3mm}
\caption{An Overview of the \Cerium{} Framework.}
\Description{}
\label{fig:Overview}
\vspace{-3mm}
\end{figure*}

\subsection{Multi-GPU FHE Requires Fine-Grained Scheduling and Optimization}

Parallelizing FHE programs across multiple GPUs can close the performance gap with ASICs. 
However, this is difficult as parallelism techniques like limb- and program-level parallelism~\cite{Cinnamon} require minimizing costly inter-GPU communication, reordering and batching of communication operations, fine-grained scheduling to overlap communication with compute, and runtime management of CUDA streams and memory buffers to yield meaningful speedups.
These tasks mandate a compiler and a runtime system that can orchestrate parallel FHE execution across-GPUs.

%% file: Sections/Design.tex
\section{Design}

\Cerium{} is a multi-GPU framework for encrypted AI based on CKKS-FHE which supports the encryption of real values.
The \Cerium{} framework consists of three components: (i) a Domain-Specific Language (DSL), (ii) an optimizing compiler, and (iii) a runtime system. 

\Cerium{}'s design philosophy is to separate application logic from system-level concerns, allowing FHE programmers to express computations without manually constructing high-performance GPU implementations. The framework automatically generates optimized GPU kernels, manages memory for terabyte-scale encrypted workloads such as Llama3-8B, and parallelizes FHE programs across multiple GPUs to narrow the performance gap with FHE ASICs.

\Cref{fig:Overview} presents a high-level view of the \Cerium{} workflow.
Programs are written using \Cerium{}’s Python embedded DSL, after which the compiler lowers them to polynomial and then limb level intermediate representations, applying optimizations at each stage. The compiler then groups limb-level IR operations into fused limb IR instructions, which are passed to the kernel code generator to produce optimized GPU kernels.
Next, the compiler constructs a kernel and memory schedule that encodes data dependencies, kernel ordering, and memory allocations. The fused kernel code is compiled into a shared library and, together with the kernel and memory schedule, is provided to the \Cerium{} runtime.
The runtime initializes the program, launches kernels according to the schedule, manages host and GPU memory, and orchestrates inter-GPU communication.
The following sections provide details on \Cerium{}'s design, outlining architectural choices, challenges encountered, and technical mechanisms developed to resolve them.

\subsection{Domain-Specific Language}
\label{sec:DSL}
\begin{figure}[ht]
\centering
\includegraphics[width=0.98\linewidth]{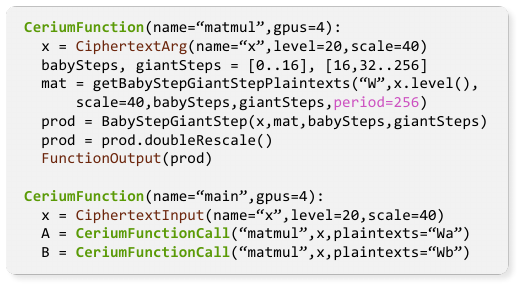}
\vspace{-4mm}
\caption{A Code Sample with the \Cerium{} DSL.}
\Description{}
\vspace{-6mm}
\label{fig:DSL}
\end{figure}

The \Cerium{} DSL is a lightweight Python-embedded DSL that enables  developers to express computations at the FHE circuit level, without dealing with low-level C++ or CUDA code. It supports static type checking to reduce programming errors and modular program decomposition through the \texttt{\Cerium{}Function} construct.
\Cref{fig:DSL} shows an example implementing matrix multiplication using the Baby-Step Giant-Step~\cite{bsgs} algorithm in the DSL. The language allows declaring plaintext types with a~\texttt{period} argument, which the compiler uses to implement the plaintext compression technique that will be introduced in~\Cref{sec:PlaintextCompression}.
It supports multi-GPU parallelization using the limb and program parallelism techniques from Cinnamon~\cite{Cinnamon}.
Crucially, the DSL hides system-level details such as memory management, threading, kernel scheduling, and cross-GPU communication, delegating them to the compiler and runtime.

\subsection{Polynomial and Limb IR Lowering}

The \Cerium{} compiler lowers programs in the DSL first to a polynomial-level IR and then to a limb-level IR, performing optimizations at each stage.
The polynomial level IR is an extension of the IR in ~\cite{Cinnamon}. Here, we make use of optimizations from prior work like hoisting~\cite{HELib}, double-hoisting~\cite{doubleHoisting}, mod down merging~\cite{MAD}, double rescale~\cite{SHARP}, limb-level parallelism~\cite{Cinnamon}. After optimization, the representation is input to the limb IR fusion passes that group instructions together for eventual kernel creation.

\subsection{Limb IR Fusion}
~\label{sec:kernel_fusion}
\begin{figure*}[ht]
\centering
\includegraphics[width=0.98\linewidth]{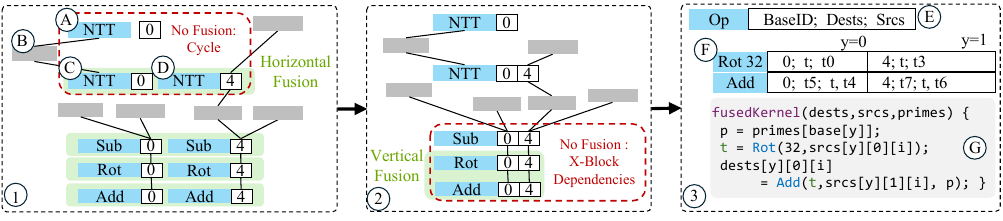}
\vspace{-2mm}
\caption{Limb IR Fusion and Code Generation}
\Description{}
\label{fig:KernelFusion}
\vspace{2mm}
\end{figure*}

The primary optimization for achieving high-performance FHE execution on GPUs is kernel fusion~\cite{100x,WarpDrive,TensorFHE}. Prior works perform this process manually, in an ad hoc manner. In contrast, \Cerium{} fully automates kernel fusion, enabling the systematic generation of high performance GPU kernels directly from FHE circuits.

The challenge in designing an automated kernel fusion technique lies in efficiently reasoning about the correctness and performance of the fused kernels.
Considering performance is necessary, as kernel fusion is not universally beneficial.
For example, kernel fusion stresses resources like register files and shared memory, and excessive fusion can lead to poor performance.
In general, automatically reasoning about kernel fusion is difficult as the search space and possibilities are very large.
\wz{What are some options for fusion? Option 1: fuse at the FHE primitive level, but lack of fusion opportunities leads to short running kernels, etc. Option 2: polynomial-level IR, very similar to limb-level, but Cerium has multi-GPU execution which requires splitting limbs across different devices. So sticking with polynomial-level IR would prevent limb-level parallelism. So then the only option is fusion at the limb-level IR, but the challenge is the size of the search space: large number of FHE limb-level IR instructions. But our insight is that...}

To simplify the search space and enable efficient reasoning, our insight is to automate fusion at the limb level. Limbs are well suited for this task because limb operations (i) constitute the atomic building blocks of RNS-CKKS computation and (ii) can be partitioned into a small set of classes that map to fundamentally different kernel structures, which are not amenable to cross-class fusion. Since the  set of limb operations and classes is small, the combinatorial possibilities of interactions between them is limited and can be explored using a rule-based approach over the limb operations.

\Cerium{} introduces Limb IR, an intermediate representation that captures fundamental limb-level operations and abstracts their kernel implementations. Limb IR instructions consist of an opcode, RNS Base ID, and destination and source operands. They provide a compact abstraction of the underlying kernels, allowing the compiler to efficiently reason about correctness, dependencies, and resource utilization.

\Cerium{}'s kernel fusion is split into two passes. \Cerium{} begins by performing horizontal fusion, i.e.,\ fusion of data-independent limb IR operations into the same kernel.
Horizontally fused instructions run in separate thread blocks of the materialized kernel. The primary benefit of this fusion is that it allows for submitting larger amounts of work to the GPU, increasing utilization and amortizing launch overheads. \circled{1} in~\Cref{fig:KernelFusion} depicts an example of horizontal fusion. It shows a DAG of unfused limb IR instructions. For brevity, we omit showing the source and destination operands as they can be inferred  from graph edges. The digits indicate the RNS Base IDs of the limb IR instruction. \rebut{For example, NTT4 indicates an NTT operation modulo the RNS prime at index 4. The gray boxes represent other limb IR instructions in the graph and are presented here to highlight dependencies between instructions.}

The \Cerium{} compiler iterates over the DAG of limb IR instructions, greedily checking for eligible horizontal fusion candidates \rebut{in a pairwise manner}. 
Limb IR instructions can be horizontally fused if they have the same opcode. For example, NTT instructions \circled{C} and \circled{D} can be fused together. However, fusion is disallowed if it results in a cycle in the graph. Thus, NTT instruction \circled{A} cannot be fused with instruction \circled{C} as fusion would create a cyclical dependency on \circled{B}. Horizontal fusion across different opcodes is disallowed as it would create control flow divergence in the generated GPU code.
\rebut{Limb IR instructions are annotated with opcode relevant metadata like level, base conversion input and output dimensions, mod down factors. \Cerium{} employs rules to ensure that the metadata factors of the two limb IR instructions that are fused with each other are the same.}

Following horizontal fusion, \Cerium{} performs vertical fusion greedily.
Vertical fusion combines instructions that contain data dependencies within the same thread, and thus improves performance by using registers to communicate data rather than expensive global memory. 
Vertical fusion requires checking for cyclical dependencies, Base ID alignment, and  cross thread block data dependencies, as sharing data across thread blocks is unsafe.
The \Cerium{} compiler reasons about cross-block dependencies using the compatibility of opcodes and operand sources. 
\wz{What are the compatibility metrics?}
\circled{2} in~\Cref{fig:KernelFusion} depicts an example of vertical fusion. Here, the (horizontally fused) rotate and add instructions are valid vertical fusion candidates as they have compatible opcodes and aligned RNS base IDs. However, the subtract instruction cannot be vertically fused with them, as this would result in both the source and destination operands of the rotate instruction residing in the same kernel, leading to unsafe cross-thread block communication from the permutation of values in the rotate instruction. 
For simplicity, \Cerium{} limits vertical fusion to elementwise operations.

Compilation efficiency requires efficiently determining whether fusion will create cyclical dependencies. 
To do so, \Cerium{} stores the parents and children of each limb IR instruction in the graph and checks whether the intersection between the parents of one instruction and the children of the other is empty, as a non-empty set implies a cycle. 
Furthermore, to keep the DAG size small, \Cerium{} performs fusion within individual \texttt{\Cerium{}Functions} defined in the DSL program. However, if a function's limb IR DAG is too large, the compiler partitions it into smaller sub-DAGs and performs fusion independently within each. 
The sub-DAG size is a configurable parameter that controls the trade-off between optimization aggressiveness and compilation latency.
We experimentally pick the default sub-DAG size.

As we show in~\Cref{sec:eval:techniques,sec:eval:CompileTimes}, \Cerium{} can compile large programs in a few minutes and its kernel fusion results in an average $2.87\times$ performance improvement over an unfused, direct polynomial to kernel baseline and a $1.21\times$ improvement over Cheddar~\cite{cheddar}.

\subsection{Kernel Code Generation}
\label{sec:CodeGeneration}

After the limb IR fusion pass, the \Cerium{} code generator lowers the abstract fused limb IR representation of the kernels into kernel code, as shown by \circled{3} in~\Cref{fig:KernelFusion}. \circled{E} depicts the format of a limb IR instruction consisting of an opcode, RNS base ID, and destination and source operands. \circled{F} shows a horizontally and vertically fused limb IR instruction. \circled{G} depicts the generated kernel (pseudo) code for \circled{F}. Here, \texttt{y} and \texttt{i} correspond to the index of horizontal fusion and index within the limb vector, respectively. 
We describe two heuristics used in \Cerium{}'s code generator to improve efficiency of generated kernels.

\subsubsection{Load Reuse.} As Elementwise kernels are bandwidth bound, we generate code that optimizes for reuse of values.
When a value is a common input to different horizontally fused limb IR instructions with the same RNS base, the \Cerium{} code generator will downgrade the horizontal fusion and place the values in the same thread. This reduces the overall bandwidth consumption, even though it comes at the cost of other factors --- occupancy, latency, and parallelism.

\subsubsection{Kernel Splitting.} 
\begin{figure}[ht]
\centering
\includegraphics[width=0.98\linewidth]{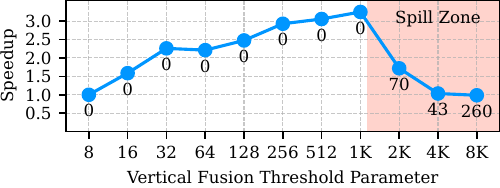}
\vspace{-3mm}
\caption{Speedup vs Vertical Fusion Threshold for a $128\times4096$  Ciphertext Matrix with a $4096\times1024$ Plaintext Matrix. Labels indicate aggregate kilobytes spilled per thread}
\Description{}
\label{fig:kernel_splitting}
\vspace{-3mm}
\end{figure}

Unrestricted, vertically fused kernels can span thousands of instructions, significantly increasing the number of operands. This places pressure on the register file and may cause costly spills.
For example, large matrix multiplication kernels contain thousands of plaintext–ciphertext multiply–accumulate operations.
~\Cref{fig:kernel_splitting} shows how the performance of ciphertext–plaintext matrix multiplication varies with the extent of vertical fusion, controlled by a threshold parameter which is a weighted sum of the count of the limb IR instructions in the kernel. Initially, increasing fusion provides substantial performance gains. However, the benefits eventually plateau as  fusing large kernels yield diminishing returns. Beyond a point, excessive fusion increases register pressure, leading to expensive spills and steep performance degradation. For instance, a parameter value of 2K results in an aggregate of 70 KB of spills per thread across all kernels, causing a $2\times$ performance drop.
To avoid this pitfall, \Cerium{} enforces a threshold on kernel size and splits kernels that exceed this limit into smaller kernels. 
\wz{I think you should explain how this threshold is determined and whether its generalizability}

Similarly, excessive horizontal fusion stresses memory capacity and bandwidth as memory locations cannot be reused across thread blocks. \Cerium{} estimates the number of memory locations required by a kernel and splits large horizontally fused kernels when required.

\subsection{Plaintext Compressed Encoding}
~\label{sec:PlaintextCompression}

As discussed in~\Cref{sec:mm_motivation}, FHE substantially increases memory requirements. This overhead is amplified by common packing schemes that repeat matrix values within plaintext vectors to reduce rotation costs~\cite{Rotom,bsgs}. The resulting repetition increases both the number and total size of the required plaintexts. This is particularly problematic for encrypted inference with models such as BERT-Base and Llama3-8B, whose large weight matrices produce plaintext storage footprints of up to $100\times$ their original size: 1.5~TB for BERT-Base and 112~TB for Llama3-8B.

\begin{figure}[ht]
\centering
\vspace{-3mm}
\includegraphics[width=0.98\linewidth]{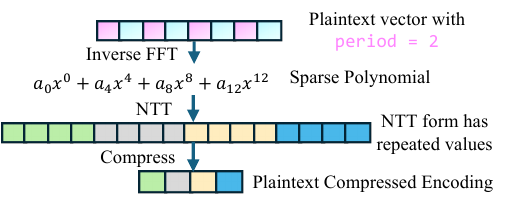}
\vspace{-2mm}
\caption{Plaintext Compressed Encoding}
\Description{}
\label{fig:plaintext_compression}
\vspace{-4mm}
\end{figure}

To conserve memory, we introduce a compression technique that exploits the algebraic interaction between power-of-2 periodicity and the CKKS encoding algorithm to reduce the memory capacity required for plaintexts. The CKKS encoding algorithm first converts a plaintext vector in the time domain to a polynomial in the frequency domain by performing an inverse FFT. As a periodic vector has fewer independent frequencies, the resulting polynomial is sparse. When this sparse polynomial is transformed back to the (bit-reversed) NTT representation, it results in contiguous, repeated values. This contiguous repetition can be compressed by storing only distinct elements and tracking their original indices. An example is shown in Figure~\ref{fig:plaintext_compression}. ~\Cref{thm:Compression} formally states this property. A full proof, along with constructions of ciphertext addition, multiplication, and matrix multiplication is available in ~\Cref{sec:appendix}.

\begin{theorem}
\label{thm:Compression}
Let $P$ be an $M$-slot vector of period $T$, where $T < M$ are powers of two.
Under the permutation $k(i)=\frac{(5^i \mod 4M)-1}{4}$, conjugate extension, IFFT, NTT, and bit-reversal, the resulting NTT representation, $P_{ntt}$ 
has a contiguous layout with at most $2T$ distinct elements.
\end{theorem}

The \Cerium{} DSL supports declaring plaintext inputs with a power of 2 \texttt{period} argument. The compiler represents them in the compressed plaintext encoding and performs the necessary indexing transformations when generating code. 
As~\Cref{sec:eval:PlaintextCompression} will show, this optimization yields dramatic memory savings of more than $100\times$, enabling running LLMs like BERT-Base and Llama3-8B in FHE.
Therefore, efficiently implementing large models in FHE requires creating packing strategies where the period is a power of two, enabling use of \Cerium{}'s compressed plaintext encoding for memory compression and performance.

\subsection{Kernel and Memory Schedule Creation}\label{sec:MemoryAndProgramSchedule}
To assemble the generated kernels into an executable program, the \Cerium{} compiler constructs a kernel and memory schedule.
This schedule contains information on the input and output operands and auxiliary data for each fused kernel. It also contains data dependencies between kernels and serves as the execution plan for the \Cerium{} runtime, which orchestrates kernel launches and manages data flow.

When creating the execution plan, intermediate ciphertext values are a significant source of memory pressure, as the total sizes of operands can exceed hundreds of GB.
To address this, \Cerium{} reuses memory locations, and performs operations in-place, conserving valuable GPU memory and eliminating costly memory management.
The challenge here is to reason about the correctness of the reuse and prevent data races; \Cerium{} does so by efficiently reasoning about kernel data access patterns using the fused limb IR representation of the kernels. 
Fused limb IR instructions begin in the static single assignment form and are topologically ordered. The compiler determines the lifetime of values and assigns address locations. If the compiler determines that a value dies in a kernel, it attempts to recycle its location. For safe reuse, the compiler only reuses memory locations within a kernel if it can ensure that the dead value is read by only one thread in the kernel. This enables the compiler to reuse its address for a subsequent write in the same thread. This is valid as instructions within a thread are always executed in order. Rotation operands are restricted from being reused within the same kernel as thread ordering across blocks is not deterministic.
Dead locations can be safely recycled across kernel boundaries as all reads are guaranteed to have completed before subsequent writes in the next  kernel. 

A second challenge is that FHE workloads exhibit heterogeneous data lifetimes. Some data persists across programs, while others change frequently. For example, evaluation keys (evalkeys) remain constant during a single inference session but differ across sessions with new keys. Plaintext weight matrices, on the other hand, are reused across inferences within the same layer but differ between layers, whereas other plaintexts, such as bootstrapping matrices or masks, are shared across both layers and inferences. Managing these distinct lifetimes efficiently requires a carefully designed memory layout and swapping strategy. 

To handle these heterogeneous workload lifetimes, the \Cerium{} compiler uses information about the type of values to create a memory layout that organizes GPU memory into dedicated memory regions for different data categories: plaintext weights, evalkeys, plaintexts, ciphertext inputs, and intermediate ciphertext values. Each pool is statically allocated and managed independently, allowing values to be updated or swapped simply by adjusting the pointer to the corresponding memory pool. This design simplifies data reuse across program executions. We revisit this design in~\Cref{sec:Runtime} and show how it plays a crucial role in enabling memory and kernel scheduling optimizations that are performed by the \Cerium{} runtime. These optimizations conserve valuable GPU memory and enable efficient host-to-GPU data movement for workloads like Llama3-8B that require more memory than is available on the GPU.

\subsection{Multi-GPU Parallelization}
~\label{sec:MultiGPU}

As FHE programs are limited by memory capacity, bandwidth, and SM count, parallelizing execution across multiple GPUs can improve performance by increasing aggregate resources. The main challenge, however, lies in inter-GPU communication overhead. Specifically, each communication operation requires synchronization across GPUs and typically lasts several microseconds, significantly impacting end-to-end performance.
To optimize FHE programs for multi-GPU execution, 
\rebut{\Cerium{} implements parallelization with two objectives:
(i) minimizing inter-GPU communication and, (ii) scheduling communication operations to avoid idling and overlap them with computation.}

\subsubsection{Minimizing Inter-GPU Communication}

\begin{figure}[ht]
\centering
\vspace{-1mm}
\includegraphics[width=0.98\linewidth]{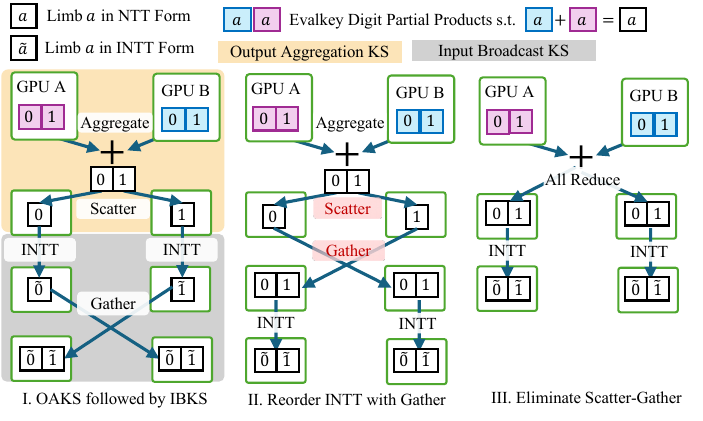}
\vspace{-4mm}
\caption{Merge aggregate scatter and all reduce across output aggregation and input broadcast keyswitches}
\label{fig:CommunicationOptimizationPass}
\Description{}
\vspace{-1mm}
\end{figure}

Cinnamon~\cite{Cinnamon} identifies limb- and program-level parallelism in FHE and proposes two key-switching schemes: input broadcast key-switching (IBKS) and output aggregation key- switching (OAKS), each with distinct communication patterns. \Cerium{} builds on this by introducing compiler optimizations to further reduce communication overhead.

IBKS broadcasts the limbs to all chips via an all-gather operation (AG) at the beginning, while OAKS performs an aggregate-and-scatter (AS) at the end. When the output of OAKS serves as an input to an IBKS, \Cerium{} fuses these steps into a single all-reduce by reordering operations, eliminating redundant scatter/gather phases.

\Cref{fig:CommunicationOptimizationPass} 
depicts this optimization.
\circled{1} shows the AS-step of an OAKS, followed by an IBKS on a 2 limb polynomial. As gather is only a data movement operation, the sequence: AS $\rightarrow$ INTT $\rightarrow$ AG can be reordered as AS $\rightarrow$ AG $\rightarrow$ INTT as shown in ~\circled{2}. 
Finally, as the scatter and gather operations cancel each other out, the communication operations can be simplified to a single all reduce (\circled{3}). 
This transformation trades minor extra computation for significantly reduced communication and is commonly found as a result of successive matrix multiplication operations as is in bootstrapping.  

Additionally, \Cerium{} removes redundant broadcasts by replacing multiple broadcast-then-accumulate sequences with a single aggregate-scatter, reducing communication calls and synchronization.

\subsubsection{Scheduling Computation and Communication} To further hide communication latency, \Cerium{} incorporates compiler transformations that reorder communication operations to occur as early as possible, allowing data transfers to proceed concurrently with downstream computation. Additionally, \Cerium{} partitions long communication operations into smaller segments, enabling finer-grained interleaving of compute and communication, thereby preventing computation from waiting for a long-running communication operation to complete.
At runtime, communication operations are dispatched to dedicated CUDA streams, distinct from those used for computation, allowing both phases to progress simultaneously.
This overlap strategy hides communication latency and ensures that available GPU resources remain fully utilized.

\subsection{\Cerium{} Runtime}
\label{sec:Runtime}

The kernels generated in~\Cref{sec:CodeGeneration} are compiled into a shared library and together with the kernel and the memory schedule from~\Cref{sec:MemoryAndProgramSchedule} are input to the \Cerium{} runtime, which manages program execution. It initializes the program, links all compiled \texttt{\Cerium{}Functions} together, sets up inputs, allocates and manages memory, and handles kernel launching. In this section, we describe some of the key features of the \Cerium{} runtime.

\subsubsection{Memory Pool Reuse}
For large programs, the intermediate values of each function consume several GB of valuable GPU memory. Allocating separate intermediate value memory pools for every function would waste dozens of GB of valuable GPU memory. Therefore, to conserve GPU memory, the \Cerium{} runtime analyzes the call graph of functions in the program and shares this memory pool between non-overlapping function calls. 

\subsubsection{CudaGraphs Generation}
\label{sec:CudaGraphs}

Kernel launch overhead is significant in FHE workloads as even small programs span thousands of kernels. \Cerium{} addresses this using CudaGraphs, an asynchronous programming model which executes a DAG of kernels with a single launch.
However, there are several challenges associated with creating CudaGraphs. 
First, the asynchronous programming model makes manually creating a CudaGraph for thousands of kernels difficult.
Second, CudaGraph creation is a significant fraction of the overall execution time and 
thus must be created offline by the runtime and 
be 
reusable across inference queries.
\Cerium{} addresses these challenges to reduce kernel launch overheads. It automatically constructs CudaGraphs by leveraging program and memory schedules to infer kernel dependencies (both true and false) and insert edges that enforce happens-before relations, preserving semantics.
While such a task is hard to do manually, the compiler and runtime-based
approach of \Cerium{} handles it relatively easily. 
When it comes to the challenges of reusing the graph across program instances and managing graph sizes for large problems, the design of the \Cerium{}'s memory layout from~\Cref{sec:MemoryAndProgramSchedule} plays an important role. To reuse graphs across programs that use different evalkeys, only the pointer to the evalkey memory pool needs to be updated, thereby avoiding the need to recreate the graph for the new program instance. Similarly, when it comes to dealing with the large graph sizes of BERT-Base and Llama3-8B, the \Cerium{} runtime creates a CudaGraph of each \Cerium{ } function and reuses these function-level graphs at every function call. Here, too, \Cerium{}'s memory pool design enables CudaGraph reuse by just requiring an update to the plaintext weight pointers.

\begin{figure}[t]
\centering
\includegraphics[width=0.93\linewidth]{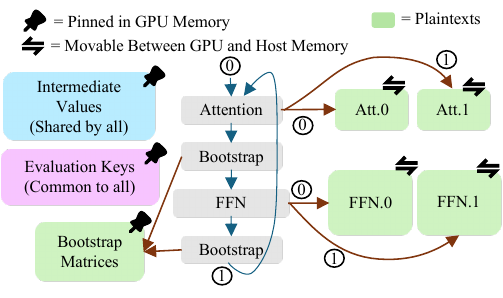}
\vspace{-3mm}
\caption{ Showing how \Cerium{}  manages memory and creates reusable CudaGraphs for a transformer inference}
\label{fig:CeriumMemoryManagement}
\Description{}
\vspace{-7mm}
\end{figure}

\subsubsection{Memory Pinning and Prefetching}
To support large LLM workloads like encrypted Llama3-8B inference that don't fit in GPU memory, \Cerium{} makes use of unified virtual memory (UVM). However, UVM is slow without guidance, and naively using UVM is detrimental to performance.
To address this challenge, the \Cerium{} runtime uses information from the heterogeneous nature of data lifetimes (Section~\ref{sec:MemoryAndProgramSchedule}) to manage host-to-device memory transfers. \Cerium{} pins frequently-accessed values---such as evalkeys and bootstrap matrices---in GPU memory and shuttles the large weight matrices between the host and GPU memory. The runtime uses information from the compiler and prefetches the weights for each layer before it is run. \Cref{fig:CeriumMemoryManagement} shows an example of this using a transformer inference. The evalkeys, intermediate values, and bootstrap matrices are all pinned in GPU memory. Before running each layer of the transformer, the runtime updates the pointer to the memory pool and prefetches it from host to GPU.

\subsection{Framework Extension}

\markBlue{
\Cerium{} supports the compilation and GPU execution of CKKS circuits and can implement any application expressible in CKKS. Its compilation pipeline is designed to be extensible and can incorporate complementary techniques that simplify FHE programming, including automatic ciphertext management, such as rescaling and bootstrapping placement~\cite{DaCapo}, and frontends that translate PyTorch or tensor programs into homomorphic circuits~\cite{Rotom,Orion}. Furthermore, as \Cerium{}'s limb IR targets RNS-based modular polynomial arithmetic, it could be extended to support other RNS-RLWE FHE schemes, including BGV~\cite{bgv} and BFV~\cite{bfv1,bfv2}.
}

%% file: Sections/Implementation.tex

%% file: Sections/Evaluation.tex
\section{Implementation}
We implement the \Cerium{} compiler in ${\approx}$25,000  lines of C++ and the runtime in ${\approx}$11,000 lines of C++/CUDA. Applications are written in Python using the \Cerium{} DSL. Inter-GPU communication uses NCCL~\cite{NCCL}.

\noindent\textbf{Compiler.} The primary compilation unit is \texttt{\Cerium{}Function}, which is intended to contain coarse-grained functionality like Matrix Multiplication, Non linear functions, Bootstrap, etc and enable modular development of the program. 
For each \texttt{\Cerium{}Function}, the compiler outputs a shared library of fused kernels, a per-device schedule file detailing kernel arguments and metadata (prime moduli, base conversion, rescale factors, etc.), information on the memory pools (evalkeys, plaintexts, intermediate values), and a call graph of functions. Limb values are formatted as [poolID:offset].

\noindent\textbf{Runtime.} During initialization, the runtime loads the \texttt{main} \texttt{\Cerium{}Function} and its callgraph. Then, it loads the information on the memory pool of each called function, allocates memory for them on the GPU, and populates them with values (plaintexts/evalkeys) from files. For the intermediate value pool, it uses the call graph of functions to calculate the maximum size of the pool, and every function is assigned an offset in this pool. This is analogous to a stack frame pointer and enables reusing the intermediate value pool across non-overlapping functions. 
The runtime then loads the schedule and shared library file for the \texttt{\Cerium{}Function}s in a leaf-to-root order and starts constructing the CudaGraph. The nodes in the CudaGraph correspond to a kernel. The runtime calculates limb value addresses and other metadata and tracks reads and writes by kernels to limb addresses to determine true and false dependencies between kernels and insert edges in the CudaGraph to enforce program semantics.
For \texttt{\Cerium{}FunctionCall}s, the runtime nests the graph of the CudaGraph of the callee function in the caller CudaGraph.

\noindent\textbf{Execution.} After constructing the CudaGraph, the runtime loads the ciphertext inputs to the program, copies them to the GPU, and launches the graph. After execution, the ciphertext outputs are copied back from the GPU to the host. This step can be repeated to execute several instances.

\section{Evaluation}
We evaluate the following aspects of \Cerium{}:
\begin{itemize}[leftmargin=10pt,topsep=2pt,parsep=1pt]
    \item Its performance on benchmarks of various sizes 
    \item Comparisons with prior GPU FHE libraries and FHE ASICs
    \item How long \Cerium{} takes to compile the benchmarks
    \item A breakdown of \Cerium{}'s optimization techniques
    \item How \Cerium{}'s memory management techniques enable and optimize encrypted LLM inference 
\end{itemize}

\begin{table*}[!ht]
\parbox{\linewidth}{
\centering
\begin{tabularx}{\linewidth}{c|Y|YYYY|Y|Y|Y}
\toprule
\multirow{2}{*}{Benchmark }  & \multirow{2}{*}{GPU }  & \multicolumn{4}{c|}{\Cerium{} Execution Time} &  Related & \multicolumn{2}{c}{\markBlue{Unencrypted}}  \\
& & $1\times$ & $2\times$ & $4\times$ & $8\times$ & Work & CPU & $1\times$ GPU  \\
\midrule
  \multirow{3}{*}{Bootstrap  } 
& B200 & 14.5~ms & 9.9~ms & 8.2~ms & 7.5~ms & C: 21~ms & \multirow{3}{*}{-} & \multirow{3}{*}{-} \\
& H100 & 19.5~ms & 13.7~ms & 10.6~ms & 9.7~ms & C: 23~ms & &  \\
& A100 & 34.2~ms &  22.5~ms & 17.9~ms & 16.5~ms & C: 40~ms & &  \\

\midrule

\multirow{3}{*}{ResNet-20 } 
& B200 & 456~ms & 340~ms & 310~ms & 298~ms & - & \multirow{3}{*}{5~ms} & 2~ms \\
& H100 & 620~ms & 464~ms & 408~ms & 386~ms & - & & 3~ms \\
& A100 & 1050~ms & 774~ms & 676~ms & 638~ms & C: 1320~ms & & 4~ms \\

\midrule

\multirow{3}{*}{BERT-Base} & B200 & 28.3~s & 19.6~s & 10.8~s & 8.8~s & - & \multirow{3}{*}{0.5~s} & 0.005~s \\

& H100 & 36.1~s & 23.91~s & 13.3~s & 10.6~s & - & & 0.007~s \\
& A100 & 66.1~s & 44.3~s & 24.6~s & 19.6~s & T: 602~s  & & 0.009~s \\

\midrule

\multirow{3}{*}{Llama3-8B} & B200  & 241~s & 147~s & 71~s & 43~s & -  & \multirow{3}{*}{1.5~s} & 0.02~s \\
& H100 & 370~s & 206~s & 92~s & 62~s &- & & 0.03~s \\

& A100 & 555~s & 351~s & 155~s & 111~s & N: 13,271~s & & 0.04~s \\

\midrule
\end{tabularx}
}
{\footnotesize C=Cheddar\footnotemark~\cite{cheddar}($1\times$), T=THOR~\cite{Thor}($1\times$A100), N=Nexus~\cite{Nexus}($4\times$A100)}
\vspace{1mm}
\caption{Benchmark Execution Times}
\vspace{-6mm}
\label{table:runtimes}
\end{table*}

\subsection{Methodology}

\textbf{Benchmarks.} We target 128-bit security using ring dimension $N=64K$, $PQ < 1782$ bits, a ternary main with hamming weight $H=32K$ and ephemeral secret with $H=32$. We use 64 28b primes as the RNS basis with $Q < 1446$ bits to implement four benchmarks of increasing sizes. 

\noindent\textbf{Bootstrapping}~\cite{bootstrappingCKKS,Lattigo} refreshes a complex vector ciphertext with a message ratio $2^{5}$ to $l=18$, consuming 33 levels. It uses a CTS and STC decomposition of 4 and 3, respectively, a 31 degree Chebyshev polynomial and a double angle number of 3~\cite{HanKi} to refresh approximately $20$ bits of precision.

\noindent\textbf{ResNet-20} implements a single encrypted classification over a $32\times32\times3$ CIFAR-10 image. We follow~\cite{ResnetFHE} to implement convolutions and approximate ReLU.
We achieve an encrypted accuracy of 91.4\%, equal to the plaintext accuracy.

\noindent\textbf{BERT} implements an encrypted BERT-Base~\cite{Bert} inference on a 128 token input. We use~\cite{Nexus} to approximate nonlinearities.
BERT-Base contains a total of 997 bootstraps.
For high accuracy, we implement softmax with max subtraction. 
We achieve an encrypted inference accuracy of 69.3\%, on the GLUE RTE dataset~\cite{GLUE,RTE}, matching the plaintext accuracy. 

\noindent\textbf{Llama3-8B} implements the prefill stage of an encrypted inference over a $128$ token prompt. \markBlue{We report the time to first token (TTFT)}. Llama3-8B contains a total of 5,890 bootstraps.
We do not use any modifications that require retraining the model. 
We validate the encrypted inference using zero-shot MCQs as 
the limited token budget prevents us from running standard LLM benchmarks.

\textbf{Configurations.} We evaluate \Cerium{} across modern datacenter systems equipped with SXM GPUs: DGX A100 (80GB), DGX H100, and DGX B200, with 1.1, 2.0, and 2.0 TB of host memory, respectively.
We parallelize workloads across 1, 2, 4 and 8 GPUs. \markBlue{The encoded model weights and evalkeys are computed offline and reside on the server.}

\subsection{Comparisons with Prior GPU FHE Work}

\footnotetext{\markBlue{For Cheddar, we ran the bootstrapping benchmark as it is open sourced. Cheddar's ResNet-20 implementation has not been released at the time of writing and the ResNet-20 results are taken from the paper.}}

\Cref{table:runtimes} reports benchmark execution time across datacenter GPUs.
\Cerium{} achieves an average $1.21\times$ speedup over Cheddar on a single A100 GPU, 
demonstrating that its automated kernel fusion, scheduling, and memory orchestration can surpass expert hand-tuned implementations.
With multi-GPU parallelization, \Cerium{} further improves performance to $2.25\times$ over Cheddar. For BERT-Base, \Cerium{} delivers a $9.12\times$ speedup, reducing inference latency from THOR's 602.2~s to 66~s on $1\times$A100. Beyond optimized kernel generation, these improvements come from \Cerium{}’s memory management techniques (Sections ~\ref{sec:PlaintextCompression},~\ref{sec:MemoryAndProgramSchedule} and ~\ref{sec:Runtime}), which allow the model to fit within GPU memory and run end-to-end. Multi-GPU scaling provides an additional $3.36\times$ speedup, reducing BERT-Base inference to 19.6~s.
For Llama3-8B, prior work does not report end-to-end encrypted inference. The closest comparison, Nexus~\cite{Nexus}, estimates performance by aggregating layer runtimes, without accounting for memory overheads, for an 8 token inference. When normalized to 128 tokens, \Cerium{} is $85.6\times$ faster.

\subsection{Comparison with FHE ASICs}
\begin{figure}[ht] 
\centering
\includegraphics[width=0.95\linewidth]{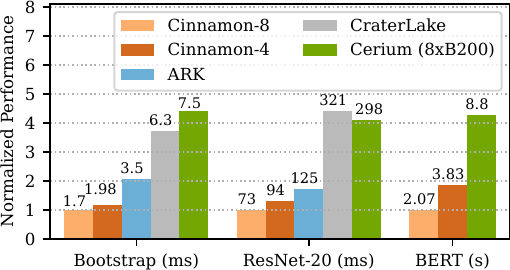}
\vspace{-3mm}
\caption{Comparing \Cerium{} to FHE ASICs
\protect\footnotemark}
\Description{}
\label{fig:results_asic}
\vspace{-3mm}
\end{figure}
\footnotetext{\markBlue{CraterLake and ARK results are omitted for BERT-Base because the published papers do not evaluate this benchmark and their implementations are not publicly available.}}

\Cref{fig:results_asic} compares \Cerium{} with the FHE ASIC designs Cinnamon~\cite{Cinnamon}, ARK~\cite{ARK}, and CraterLake~\cite{Craterlake}, with execution times normalized to Cinnamon-8. The ASIC baselines are derived from simulation studies rather than silicon measurements; therefore, the comparison reflects projected ASIC performance.

Using $8\times$ B200 GPUs, \Cerium{} achieves average speedups of $1.06\times$, $2.33\times$, $3.03\times$, and $4.40\times$ over CraterLake, ARK, Cinnamon-4, and Cinnamon-8, respectively. These results show that a highly optimized software stack can make commercial GPU hardware competitive with specialized FHE accelerators. To the best of our knowledge, \Cerium{} is the first system to demonstrate sub-$10$\,ms bootstrapping on commercial hardware, achieving $7.5$~ms, as well as encrypted BERT-Base inference in under $10$~s, achieving $8.8$~s.

\subsection{Compile Time and CudaGraph Creation Time}
~\label{sec:eval:CompileTimes}
\begin{table}[!htbp]
\centering
\begin{tabularx}{\linewidth}{Y|Y|Y}
\toprule
Benchmark  & \Cerium{} Compile Time &  CudaGraph Creation Time \\
\midrule
Bootstrap 
& 5.5 s & 350 ms \\
ResNet-20 
& 13.9 s & 1.75 s \\
BERT-Base 
& 2 min 47 s &  4.75 s \\
Llama3-8B 
& 11 min 29 s  & 24.45 s\\
\bottomrule
\end{tabularx}
\caption{Compilation and CudaGraph Creation Times}
\label{table:CompileTimes}
\vspace{-6mm}
\end{table}

~\Cref{table:CompileTimes} reports the compilation time of the \Cerium{} framework and the corresponding CudaGraph creation time across our benchmarks. \Cerium{} compiles small programs such as Bootstrap and ResNet-20 within seconds (5.5 s and 13.9 s, respectively), and scales efficiently to large encrypted workloads, compiling BERT-Base and Llama3-8B in 2 min 47 s and 11 min 29 s, respectively. The runtime’s CudaGraph creation incurs minimal overhead, completing in under 30 s even for Llama3-8B. In contrast, manually implementing and optimizing these kernels and graphs would require several hours or days of expert GPU and FHE programming effort. \markBlue{As CudaGraph creation is a significant portion of overall runtime, it must be amortizable across several inferences. \Cerium{}'s memory pool architecture enables the same CudaGraph to be reused for several successive inferences at the cost of a few pointer updates to the relevant pools (\Cref{sec:CudaGraphs}).}
These results demonstrate that \Cerium{} can rapidly generate optimized, end-to-end GPU implementations from circuit-level FHE descriptions, significantly accelerating development cycles and enabling non-expert programmers to efficiently deploy complex encrypted workloads.

\subsection{\Cerium{} Techniques Evaluation}
~\label{sec:eval:techniques}
In this section, we evaluate the techniques proposed in this paper. 
All experiments are performed on the $1\times$B200 GPU configuration, except for the multi-GPU evaluation.

\subsubsection{\textbf{Horizontal and Vertical Fusion}} 

\begin{figure}[ht] 
    \centering
    \hspace{-3mm}
    \begin{subfigure}[h]{0.57\linewidth}
        \centering        \includegraphics[width=\textwidth]{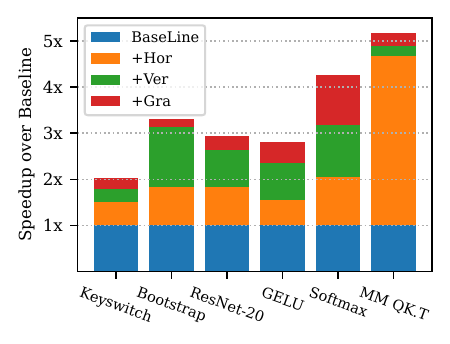}
          \vspace{-6mm}
        \label{fig:KernelFusionEval}
    \end{subfigure}%
    \begin{subfigure}[h]{0.21\linewidth}
        \centering
          \includegraphics[width=\textwidth]{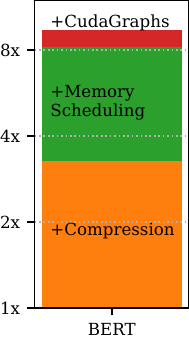}
    \end{subfigure}
    \begin{subfigure}[h]{0.21\linewidth}
        \centering
          \includegraphics[width=\textwidth]{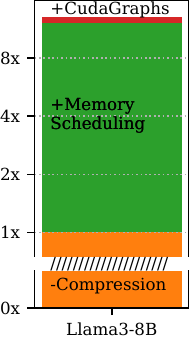}
    \end{subfigure}
    \hspace{-3mm}
    \hfill
    \vspace{-1mm}
    \caption{\Cerium{} Techniques: (a) Kernel Optimization;  Memory Optimizations for (b) BERT-Base, (c)  Llama3-8B}
    \Description{}
    \vspace{-1mm}
\end{figure}

To assess \Cerium{}’s horizontal and vertical fusion passes, we evaluate six workloads: keyswitching ($l=48,d=4)$, bootstrapping, ResNet-20, GELU, softmax and Ciphertext-Ciphertext Matmul (MM QK.T). 
GELU, softmax and MM QK.T are taken from the BERT-Base benchmark; for softmax, we report performance excluding internal bootstrapping time. The baseline is an unfused implementation where each polynomial operation is lowered to a separate kernel.
For bootstrapping and ResNet-20, horizontal fusion provides speedups of 
$1.98\times$ and 
$1.84\times$, respectively. Vertical fusion adds another 
$1.68\times$ and 
$1.43\times$, resulting in overall improvements of 
$3.32\times$ for bootstrapping and 
$2.63\times$ for ResNet-20.
For GELU, the unfused implementation takes 84.98 ms. Horizontal fusion reduces the runtime to 54.87 ms ($1.54\times$), and vertical fusion reduces it further to 36.13 ms ($1.51\times$ over horizontal).
For softmax, the unfused version takes 235.4 ms. Horizontal fusion lowers this to 115.11 ms ($2.04\times$), and vertical fusion further reduces the time to 74.09 ms ($1.55\times$ over horizontal).
MM QK.T (ciphertext multiplication between 128x128 matrices) shows a $4.8\times$ speedup (195 ms to 40.8 ms) from Horizontal fusion as it is a very wide benchmark with a small depth.
In all cases, the speedup is significantly larger than the speedup of a single keyswitch, demonstrating the importance of fusion across large sections of the program and the versatility of \Cerium{}'s compiler-driven approach.

\subsubsection{\textbf{Automatic Fusion}}
\begin{figure}[ht]
\vspace{-2mm}
\centering
\includegraphics[width=0.99\linewidth]{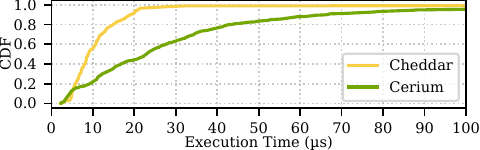}
\vspace{-3mm}
\caption{CDF for Bootstrapping Kernels}
\Description{}
\label{fig:kernel_cdf}
\vspace{-2mm}
\end{figure}

\begin{figure*}[ht] 
    \centering
    \hspace{-6mm}
    \begin{subfigure}[h]{0.3\linewidth}
        \centering
\includegraphics[height=1.2in]{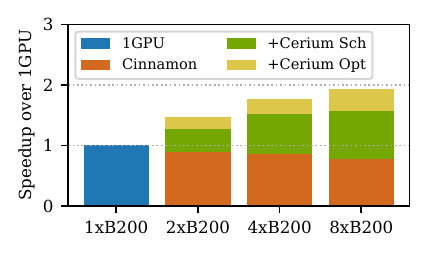}
        \vspace{-4mm}
    \end{subfigure}
    \hspace{-2mm}
    \begin{subfigure}[h]{0.69\linewidth}
        \centering 
        \includegraphics[height=1.1in]{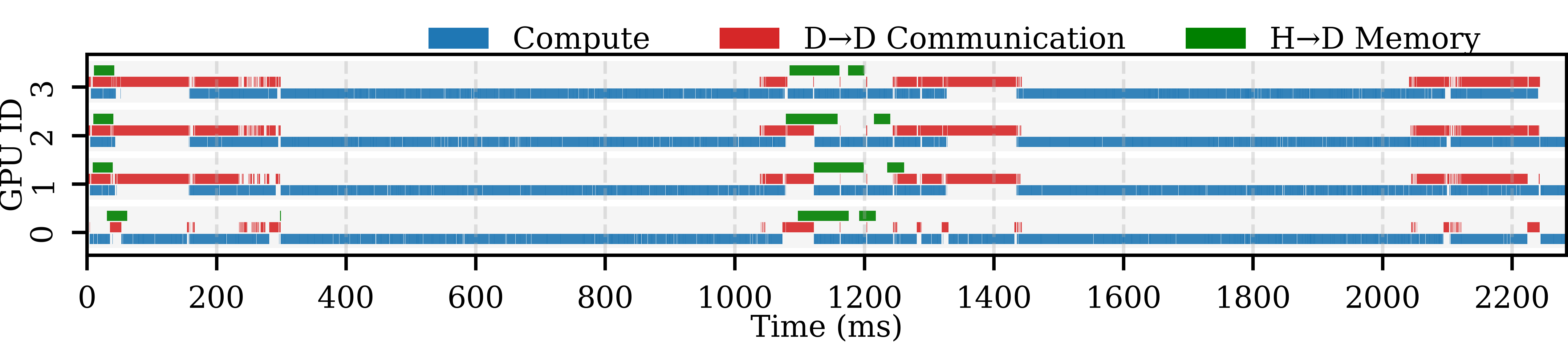}
        \vspace{-6mm}
    \end{subfigure}
    \caption{(a) \Cerium{}'s Multi-GPU Optimizations; (b) Llama3-8B Transformer Block Execution Timeline}
    \label{fig:MultiGPU}
    \label{fig:LlamaTimeline}

    \Description{}
    \vspace{-2mm}
\end{figure*}

To evaluate the benefits of \Cerium{}’s automated kernel fusion, we compare its compiler-generated kernels with Cheddar’s~\cite{cheddar} manually fused kernels. \Cerium{} reduces the number of kernels required for bootstrapping from 1490 in Cheddar to 472, a $3.1\times$ reduction.
~\Cref{fig:kernel_cdf} presents the CDF of kernel execution times for both systems on bootstrapping. In Cheddar, 56.1\% of kernels execute in under 10 µs and 97.5\% complete within 25 µs, indicating a large number of short-running kernels. This reflects the limitations of manual fusion, where identifying opportunities across larger program regions is difficult. For instance, Cheddar doesn't support fusion across the two chebyshev polynomials. 
In contrast, \Cerium{} discovers fusion opportunities across the entire program to generate larger kernels, with 21.7\% executing under 10 µs and 55.4\% under 25 µs. Consequently, \Cerium{} achieves higher average SM active cycles of $94.1\%$, compared to Cheddar’s $72.1\%$, resulting in a $22\%$ speedup.

\subsubsection{\textbf{CudaGraphs}}
To mitigate kernel launch overheads, the \Cerium{} runtime uses CudaGraphs to specify the computation as an asynchronous graph of kernels that can be executed with a single low-overhead launch. CudaGraphs yield consistent performance gains across all benchmarks: a $7\%$ speedup for bootstrapping, a $12\%$ speedup for ResNet-20, a $14\%$ speedup for BERT-Base, and a $10\%$ speedup for Llama3-8B, resulting in an average improvement of $11\%$.

\subsubsection{\textbf{Plaintext Compressed Encoding}} 
\label{sec:eval:PlaintextCompression} Packing plaintext vectors with power-of-two periods enables the \Cerium{} compiler to compress them without changing the computation. This optimization has two major benefits. First, it dramatically shrinks the memory footprint of encoded model weights, making previously intractable models feasible to run under FHE. For BERT-Base, this technique achieves a $96\times$ reduction, compressing 1.5 TB of encoded weights down to 16.6 GB --- small enough to fit entirely in GPU memory. Second, compression reduces pressure on memory bandwidth. Collectively, compression improves performance by $3.25\times$ from 266.4 s to 81.68 s. The effect is even more pronounced for larger models. For Llama3-8B, plaintext compression reduces the encoded weight footprint by $119\times$ from 112 TB to 982 GB. Without this reduction, the model would exceed GPU and host memory capacity by two orders of magnitude, making end-to-end FHE inference entirely infeasible. These results highlight how \Cerium{}'s compiler exploits structural properties in plaintext encodings to unlock substantial practical scalability in FHE workloads.

\subsubsection{\textbf{Memory Schedule}}

To avoid the overhead of repeatedly allocating and freeing intermediate ciphertexts, \Cerium{} constructs a kernel and memory schedule that captures data dependencies across kernels. The runtime uses this schedule to reuse memory locations and eliminate online allocation. We evaluate this approach against a baseline that allocates and frees memory for every intermediate value. Online memory management slows bootstrapping from 15.1 ms to 35.57 ms, a $2.3\times$ slowdown. For ResNet-20, the slowdown is $2.36\times$ (509 ms to 1206 ms).
For BERT, reusing memory locations leads to a further speedup of $2.51\times$ from 81.68 to 32.55 seconds. These results demonstrate the importance of \Cerium{}’s compiler and runtime directed memory reuse, which eliminates the high cost of online memory operations.

Llama3-8B has a working set size that far exceeds the capacity of GPUs. \Cerium{} addresses this challenge by using the memory layout, pinning, and prefetching techniques described in~\Cref{sec:MemoryAndProgramSchedule,sec:Runtime} to coordinate data movement between host and GPU. When evaluated against a baseline that uses UVM for all allocations without prefetching,
\Cerium{}'s optimizations yield a $12.1\times$ speedup for Llama3-8B, reducing execution time from 3334 to 275 seconds and highlighting the importance of a memory management system.

\subsubsection{\textbf{Multi GPU Parallelization and Scheduling}}
We evaluate multi-GPU parallelized bootstrapping using B200 GPUs. Our baselines are a single-GPU baseline and a basic multi-GPU implementation using Cinnamon's~\cite{Cinnamon} parallel keyswitching algorithm in a single stream. On average, the baseline implementation performs $1.2\times$ slower than a single GPU's 14.5 ms. However, when enabling \Cerium{}'s multi-GPU scheduling techniques to reorder and overlap compute and communication operations, performance improves on average to $1.45\times$ the single GPU case.
Furthermore, using \Cerium{}'s new communication minimization passes results in a $44\%$ reduction in bytes communicated over Cinnamon and improves performance by an additional $1.18\times$.

\Cref{fig:LlamaTimeline}(b) shows the timeline of a Llama3-8B transformer block across four B200 GPUs. \Cerium{} overlaps computation with device-to-device communication and host-to-device transfers, hiding $99\%$ of host-to-device transfer latency and $58\%$ of device-to-device communication latency.

Our evaluation demonstrates that in order to accelerate FHE over multiple GPUs and close the performance gap with FHE ASICs, the algorithmic techniques from Cinnamon alone are not enough and they require a co-designed compiler and runtime system to deliver meaningful speedups.

%% file: Sections/RelatedWork.tex
\section{Other Related Work}
CROSS~\cite{Cross} targets FHE primitive acceleration on TPUs, but does not run end-to-end applications. \Cerium{} in contrast, runs end-to-end applications up to Llama3-8B.  GME~\cite{GME} introduces 186.2 $\text{mm}^2$ of additional logic to an AMD MI100 to support FHE yet trails ASICs by more than $10\times$. In contrast, \Cerium{} approaches within $3\times$ of FHE ASIC performance without any hardware modifications, relying solely on software optimizations. BOLT~\cite{Bolt} combines HE and MPC for transformer inference, but its online MPC communication costs dominate: even on a LAN, 128-token BERT-Base inference requires 91 s. \Cerium{} shows that the same workload can now run fully in FHE on GPUs in 8.8 s, avoiding the networking bottleneck of MPC.
Concurrent efforts~\cite{Nexus,Thor,FHEFriendlyLLM} explore FHE transformer inference, but progress has been constrained by the absence of a general, high-performance GPU framework. By automatically generating optimized GPU implementations, \Cerium{} aims to unlock faster research and deployment of FHE-based applications.

Gouert et~al.~\cite{gouert2025hardware} target Boolean-gate FHE, showing multi-GPU scaling for small non-ML benchmarks.
MAGE~\cite{Mage} exploits the static nature of memory access patterns of encrypted computation to optimize disk-memory transfers. 
Finally, we remark that as FHE ASICs~\cite{Craterlake,ARK,Cinnamon} are statically scheduled, they cannot support workloads like Llama3-8B that exceed accelerator memory and require host-to-accelerator memory orchestration, highlighting a major scalability limitation.

\section{Potential Use Cases for FHE}

\markBlue{
FHE is particularly compelling when the value of confidentiality exceeds the performance overheads.
This holds, for instance, when organizations hold valuable data that cannot be outsourced in plaintext due to confidentiality, intellectual property, or regulatory concerns.
}
\markBlue{
\Cerium{} demonstrates FHE-based encrypted inference for transformer models at the scale of BERT-Base and Llama3-8B, with processing times on the order of tens of seconds. 
While this latency is unsuitable for most interactive services, it is practical for offline analysis, asynchronous review, and batch processing, 
where protecting sensitive data can matter more than quick response times.
We illustrate a few such use cases. 
}

\markBlue{
Consider a hospital running encrypted inference on a genomic-risk or disease-prediction model. The result may inform a clinical decision or prioritize further review, both of which can tolerate several seconds of latency. 
However, inputs such as genetic data, medical images, and patient records may be too sensitive to disclose to an external model provider~\cite{heOncology,heprs,hefa}.
Similarly, a law firm could use encrypted inference during legal discovery to identify relevant documents within a client's internal corpus without needing to view the entire corpus~\cite{akavia2019search}.
Here, tens of seconds of processing time is acceptable relative to a discovery process that typically spans weeks.
}

\markBlue{
Similar situations arise in financial analysis and software security. A financial institution could run an anti-money-laundering or credit-risk model over encrypted customer transaction data, where a delay of dozens of seconds is minor compared with the consequences of exposing sensitive financial records~\cite{nistFHE,oecd2025pets}.
Likewise, an organization could submit encrypted proprietary source code to an AI-assisted vulnerability-detection service, enabling the service to identify potential vulnerabilities in the code without revealing it to the service provider~\cite{kim2026ppvul}.
As vulnerability analysis can be performed offline, several seconds of latency is acceptable.
}


\section{Future Work}

\markBlue{
\Cerium{} demonstrates encrypted transformer inference at the scale of BERT-Base and Llama3-8B, but several challenges remain to enable broader deployment. One important direction is efficient encrypted KV-cache management for autoregressive decoding, including the design of suitable packing strategies and the management of the capacity and data-movement costs associated with subsequent-token generation. Another is studying the tradeoff between model accuracy and the computational cost of polynomial approximations for nonlinear operations, including robustness to adversarial and out-of-distribution inputs. \Cerium{} provides a framework for evaluating the performance of alternative packing strategies and polynomial approximations alongside their effects on model accuracy.
}

\markBlue{
Future work could also improve energy efficiency through hardware- and compiler-level techniques that reduce the cost of encrypted arithmetic, communication, and memory movement. Finally, continuous batching of encrypted requests could improve GPU utilization and system throughput, although it requires careful management of the substantial memory footprint of concurrent FHE computations. These directions could build on \Cerium{}'s limb-level IR, kernel-fusion passes, and memory-pool design.
}

%% file: Sections/Conclusion.tex
\section{Conclusion}

FHE usage has been constrained by the assumption that commodity hardware is insufficient and custom silicon is required. 
By building a highly optimized software stack for GPUs, \Cerium{} is able to match the performance of FHE ASICs -- showing that the performance gap between commodity GPUs and custom silicon can be bridged entirely in software and without building custom hardware. The \Cerium{} framework automates the creation of optimized multi-GPU FHE programs using compiler techniques for kernel generation, optimization, and scheduling. \Cerium{} introduces techniques that exploit algebraic structure to reduce the memory capacity requirements of encrypted inference by more than $100\times$, minimize the memory overhead of FHE programs, and enable running terabyte-scale encrypted inference workloads for the first time.

%% file: Sections/Acknowledgements.tex
\begin{acks}
We would like to thank the anonymous reviewers and shepherd for feedback on the paper; Edward Chen for comments, feedback and help drafting the manuscript.
This research was funded by NSF Grants CNS-2238671 and CCF 2217016, an Amazon Web Services Grant, and a CMU Cylab Grant. 
\end{acks}

%% file: Sections/Appendix.tex
\newpage
\appendix

\textit{Note: This appendix contains supplementary material that has not been peer-reviewed.}

\section{Plaintext Compressed Encoding}
\label{sec:appendix}
\subsection{Proof of Periodic Compression}

Let \( V \in \mathbb{C}^M \) be \(T\)-periodic, where $T<M$ are powers of 2. 
Let $y(X) \in \mathbb{C}[X]/(X^{N} + 1)$
denote its CKKS encoding, for $N=2M$. We want to show that the periodicity of \( V \) induces: 
1) coefficient sparsity in $y$, with support restricted to indices spaced by $\frac{N}{2T}$, and
2) periodic structure in the NTT domain, with evaluation values repeating every $\frac{N}{2T}$ slots.

\begin{lem}
\label{lem:permutation}
Let $X$ be a permutation of $V$ with index map of $k(i) \;:=\;\frac{(5^i \bmod 4M)-1}{4}, 0\leq i < M$. If $V$ is periodic with period $T$, then $X$ is also periodic with period $T$.
\end{lem}

\begin{proof}
First, $k(i)$ is a valid permutation as $5^i \pmod{4M}$ generates $M$ unique odd integers of the form $4k+1$ for $k \in [0, M-1)$.
Given index $i_l \in [0, M)$, let  $k_l = k(i_l)$.
WLOG, let $i_2 = i_1 + \lambda T$, for $\lambda\in\mathbb{Z}$. Then, 
$k_2 - k_1 = (5^{i_2} - 5^{i_1})/4
=5^{i_1}((1+4)^{\lambda T} - 1)/4$.
Applying the Binomial Theorem, we get 
$k_2 - k_1 = 5^{i_1} \left( \left(\sum_{j=0}^{\lambda T} \binom{\lambda T}{j}4^j\right) - 1\right)/4
= 5^{i_1} \cdot 4T \textcolor{blue}{\left( \lambda + \frac{\lambda(\lambda T - 1)}{2} \cdot 4 + ... \right)}/4$.
Let the integer term in the \textcolor{blue}{blue} parentheses be $m$.
Then $k_2 - k_1 = 5^{i_1} \cdot T m \equiv 0 \pmod T$.
Since the permutation is bijective, different equivalence classes in $X$ cannot fall into the same equivalence class in $V$.
\end{proof}

Let $Y$ be the conjugate part extension vector of $X$, formally defined as: $Y[2k] = X[k] \quad \text{and} \quad Y[2(M-k)-1] = \overline{X[k]}$.
By definition, $Y$ is of size $N = 2M$ with a period $2T$.

\begin{lem}[Sparsity of IFFT on $Y$]
\label{lem:ifft}
Let $N$ and $2T$ be powers of two such that $2T \mid N$. 
Let $\omega$ be a primitive $2N$-th root of unity.
Let $y$ be the IFFT of $Y$, where $y[\hat{j}] = \frac{1}{N} \sum_{j=0}^{N-1} Y[j]\;\omega^{-\hat{j}(2j+1)}$.
If $Y$ is periodic with period $2T$, then $y$ is a $2T$-sparse polynomial.
\end{lem}

\begin{proof}
We can rewrite the IFFT summation by breaking it into two nested summations: 
$y[\hat{j}] = \frac{1}{N} \sum_{t=0}^{2T-1} \sum_{s=0}^{(N/2T)-1} Y[t + s(2T)] \omega^{-\hat{j}(2(t + s(2T)) + 1)}$.
Since $Y$ is periodic with period $2T$, $Y[t] = Y[t + s(2T)]$.
Thus: 
$$y[\hat{j}] = \frac{1}{N} \sum_{t=0}^{2T-1} Y[t] \omega^{-\hat{j}(2t+1)} \textcolor{blue}{\left(\sum_{s=0}^{(N/2T)-1} (\omega^{-4T\hat{j}})^{s}\right)}$$
Using the orthogonality properties of the roots of unity, we can conclude that the inner summation (in blue) is non-zero only when $\hat{j} \equiv 0 (\bmod \  N/2T)$; and thus $y$ is a $2T$-sparse polynomial.  
\end{proof}

Moreover, for a sparse vector $y$, scaling, rounding, and integer modulus preserve sparsity.

\begin{lem}[Periodicity of NTT on $y$]
\label{lem:periodntt}
Let $N$ and $2T$ be powers of two such that $2T \mid N$. 
For polynomial $y$ of degree $N$, let $V_{ntt}$, also of size $N$, be the set of polynomial evaluations at the roots of unity $\omega$. 
If $y$ is structured $2T$-sparse, i.e.,
$y[j] = 0\,\text{ unless }\, 
j \equiv 0 \bmod{\frac{N}{2T}}$,
then $V_{ntt}$ is periodic with period $2T$.
\end{lem}

\begin{proof}
Use the property that $\omega^{(2j+1)N/2T}$ is $2T$-periodic.
\end{proof}

\begin{lem}[Contiguity of Bit Reversal on Periodic Vectors]
\label{lem:bitrev}
Let $N$ and $2T$ be powers of 2, and $V_{ntt}$ be a periodic vector of size $N$ and $2T$. Then a bit reversal transformation on $V_{ntt}$ results in a contiguous vector.
\end{lem}

\begin{proof}
By periodicity, $i_1 \equiv i_2 \pmod {2T} \implies V_{ntt}[i_1] = V_{ntt}[i_2]$.
Also, since $N$ and $2T$ are powers of 2, lower $log_2(2T)$ bits of $i_1$ and $i_2$ are equal. Thus after bit reversal, the upper $log_2(2T)$ bits of $i_1$ and $i_2$ will be the same, forming a contiguous vector.
\end{proof}

Therefore, given a periodic plaintext vector $V$, Lemma~\ref{lem:permutation} shows the preservation of periodicity after permutation. Then applying Lemmas~\ref{lem:ifft},~\ref{lem:periodntt} show the preservation of periodicity (albeit doubled) throughout the CKKS encoding. Finally, Lemma~\ref{lem:bitrev} shows that bit reversal on the periodic NTT representation creates a contiguous vector, which sets up the vector for compression.

\subsection{Homomorphic Operations over Compressed Plaintext Encodings}

Note that the NTT representation is used for efficient ciphertext multiplication. ~\Cref{fig:algo:ctPtMul} shows how \Cerium{} implements ciphertext-plaintext multiplication when the plaintext limb vector is compressed. Homorphic addition can be implemented similarly.

\begin{figure}[h]
\lstset{language=Python} 
\begin{lstlisting}[frame=single,escapeinside={|*}{*|},commentstyle=\color{black}]
# Inputs: Ciphertext |*$C$*| = |*$(C_0,C_1)$*| in RNS-NTT form
#  Plaintext |*$P_{com}$*| with period |*$T$*| in Compressed Form
def Ct_CompPt_Mul(|*$C$*|,|*$P_{com}$*|,|*$T$*|):   
  for j,i in (|*$C$*|.num_limbs |*$\times$*| N):
     |*$D_0$*|[j][i] = |*$C_0$*|[j][i] * |*$P_{com}$*|[j][i >> log2(M/T)]
     |*$D_1$*|[j][i] = |*$C_1$*|[j][i] * |*$P_{com}$*|[j][i >> log2(M/T)]
  return |*$(D_0,D_1)$*|
\end{lstlisting}
\vspace{-3mm}
\caption{Ciphertext - Compressed Plaintext Multiplication}
\Description{}
\label{fig:algo:ctPtMul}
\end{figure}

~\label{sec:matmul}
\begin{figure*}[ht]
\centering
\hspace{-3mm}
\includegraphics[width=0.98\linewidth]{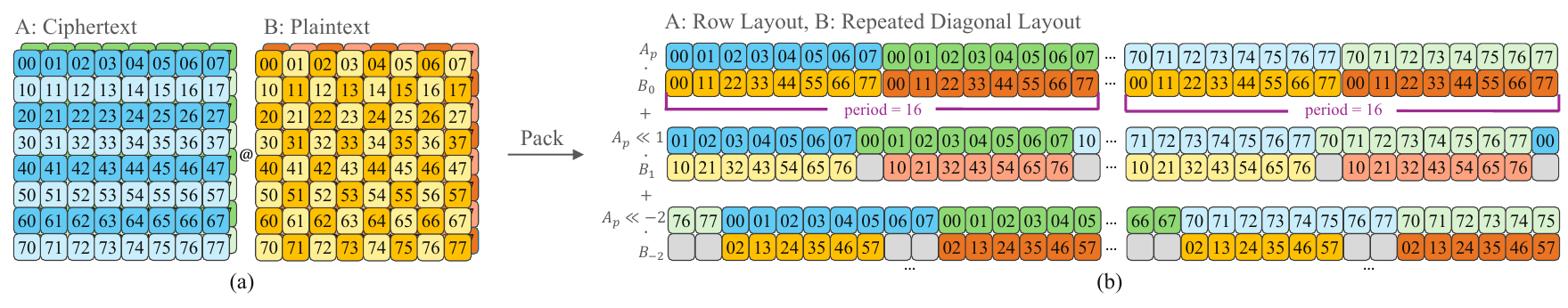}
\hspace{-3mm}
\vspace{-3mm}
\caption{\rebut{Matrix Multiplication with Periodic Plaintexts}}
\Description{}
\label{fig:MatMul}
\end{figure*}
The algorithm in~\Cref{fig:algo:ctPtMul} can be extended to matrix multiplication when plaintext matrices are carefully packed into vectors with powers of two periods.
~\Cref{fig:MatMul} describes ciphertext-compressed plaintext multiplication with the help of an example where two tensors $A$ and $B$ of size $(2,n=8,n=8)$ are multiplied using $M= 2n^2 = 128$ slots. The operation performs batched multiplication of the two  $n \times n$ tiles. Here, $A$ is the ciphertext and $B$ is the plaintext. We pack both tiles of $A$ interleaved row-wise as shown in $A_p$. $B$ is packed in a periodic diagonal form. Plaintext $B_i$ contains periodic packing of the $i^{th}$ lower diagonal of both tiles of $B$ with no wrap around. Negative values of $i$ correspond to upper diagonals. We zero pad lower diagonals at the end and upper diagonals at the start. The matmul can be written as $ \sum_{i=-n}^{n}(A_p << i) \cdot B_i$ and the result is row-packed, similar to A. The rotations required can be optimized using BSGS to $2\sqrt{2n} = 8$ rotations. This method requires $2n - 1 = 15$ plaintext vectors, each of which is periodic with $T = 2n = 16$. Thus, they can be compressed by a factor $M/T = n = 8$, resulting in an effective plaintext count of $2 - 1/n$. Larger matrices can be padded and broken up into similar tiles to implement matrix multiplication.

\subsection{Generality of Power of 2 Periodic Packings}
As most tensor operations in FHE and AI are performed over tensors sizes that are multiples of powers of 2, \Cerium{}'s plaintext compression is generalizable to most workloads. In case tensors are not powers of 2, they can be padded to a multiple of a power of 2 to implement the compression.

%% file: refs.bib
@misc{bootstrappingCKKS,
      author = {Jung Hee Cheon and Kyoohyung Han and Andrey Kim and Miran Kim and Yongsoo Song},
      title = {Bootstrapping for Approximate Homomorphic Encryption},
      howpublished = {Cryptology {ePrint} Archive, Paper 2018/153},
      year = {2018},
      url = {https://eprint.iacr.org/2018/153}
}

@misc{bsgs,
      author = {Jung Hee Cheon and Kyoohyung Han and Andrey Kim and Miran Kim and Yongsoo Song},
      title = {Bootstrapping for Approximate Homomorphic Encryption},
      howpublished = {Cryptology {ePrint} Archive, Paper 2018/153},
      year = {2018},
      url = {https://eprint.iacr.org/2018/153}
}

@inproceedings{Craterlake,
author = {Samardzic, Nikola and Feldmann, Axel and Krastev, Aleksandar and Manohar, Nathan and Genise, Nicholas and Devadas, Srinivas and Eldefrawy, Karim and Peikert, Chris and Sanchez, Daniel},
title = {{CraterLake}: A Hardware Accelerator for Efficient Unbounded Computation on Encrypted Data},
year = {2022},
isbn = {9781450386104},
publisher = {Association for Computing Machinery},
address = {New York, NY, USA},
url = {https://doi.org/10.1145/3470496.3527393}, doi = {10.1145/3470496.3527393}, booktitle = {Proceedings of the 49th Annual International Symposium on Computer Architecture}, pages = {173–187}, numpages = {15}, location = {New York, New York}, series = {ISCA '22} }

@inproceedings{Sharp, author = {Kim, Jongmin and Kim, Sangpyo and Choi, Jaewan and Park, Jaiyoung and Kim, Donghwan and Ahn, Jung Ho}, title = {SHARP: A Short-Word Hierarchical Accelerator for Robust and Practical Fully Homomorphic Encryption}, year = {2023}, isbn = {9798400700958}, publisher = {Association for Computing Machinery}, address = {New York, NY, USA}, url = {https://doi.org/10.1145/3579371.3589053}, doi = {10.1145/3579371.3589053}, booktitle = {Proceedings of the 50th Annual International Symposium on Computer Architecture}, articleno = {18}, numpages = {15}, location = {Orlando, FL, USA}, series = {ISCA '23} }

@INPROCEEDINGS{ARK,
  author={Kim, Jongmin and Lee, Gwangho and Kim, Sangpyo and Sohn, Gina and Rhu, Minsoo and Kim, John and Ahn, Jung Ho},
  booktitle={2022 55th IEEE/ACM International Symposium on Microarchitecture (MICRO)}, 
  title={{ARK}: Fully Homomorphic Encryption Accelerator with Runtime Data Generation and Inter-Operation Key Reuse}, 
  year={2022},
  volume={},
  number={},
  pages={1237-1254},
  doi={10.1109/MICRO56248.2022.00086}}

@inproceedings{ckks,
  title={Homomorphic encryption for arithmetic of approximate numbers},
  author={Cheon, Jung Hee and Kim, Andrey and Kim, Miran and Song, Yongsoo},
  booktitle={Advances in Cryptology--ASIACRYPT 2017: 23rd International Conference on the Theory and Applications of Cryptology and Information Security, Hong Kong, China, December 3-7, 2017, Proceedings, Part I 23},
  pages={409--437},
  year={2017},
  organization={Springer}
}

@inproceedings{bgv,
author = {Brakerski, Zvika and Gentry, Craig and Vaikuntanathan, Vinod},
title = {(Leveled) Fully Homomorphic Encryption without Bootstrapping},
year = {2012},
isbn = {9781450311151},
publisher = {Association for Computing Machinery},
address = {New York, NY, USA},
url = {https://doi.org/10.1145/2090236.2090262},
doi = {10.1145/2090236.2090262},
booktitle = {Proceedings of the 3rd Innovations in Theoretical Computer Science Conference},
pages = {309–325},
numpages = {17},
location = {Cambridge, Massachusetts},
series = {ITCS '12}
}

@misc{Nexus,
    author = {Jiawen Zhang and Jian Liu and Xinpeng Yang and Yinghao Wang and Kejia Chen and Xiaoyang Hou and Kui Ren and Xiaohu Yang},
    title = {Secure Transformer Inference Made Non-interactive},
    howpublished = {Cryptology ePrint Archive, Paper 2024/136},
    year = {2024},
    url = {https://eprint.iacr.org/2024/136}
    }

@article{ResnetFHE,
  author       = {Joon{-}Woo Lee and
                  HyungChul Kang and
                  Yongwoo Lee and
                  Woosuk Choi and
                  Jieun Eom and
                  Maxim Deryabin and
                  Eunsang Lee and
                  Junghyun Lee and
                  Donghoon Yoo and
                  Young{-}Sik Kim and
                  Jong{-}Seon No},
  title        = {Privacy-Preserving Machine Learning with Fully Homomorphic Encryption
                  for Deep Neural Network},
  journal      = {CoRR},
  volume       = {abs/2106.07229},
  year         = {2021},
  url          = {https://arxiv.org/abs/2106.07229},
  eprinttype    = {arXiv},
  eprint       = {2106.07229},
  bibsource    = {dblp computer science bibliography, https://dblp.org}
}

@misc{HanKi,
      author = {Kyoohyung Han and Dohyeong Ki},
      title = {Better Bootstrapping for Approximate Homomorphic Encryption},
      howpublished = {Cryptology ePrint Archive, Paper 2019/688},
      year = {2019},
      url = {https://eprint.iacr.org/2019/688}
}

@InProceedings{CryptoNets,
  title = 	 {CryptoNets: Applying Neural Networks to Encrypted Data with High Throughput and Accuracy},
  author = 	 {Gilad-Bachrach, Ran and Dowlin, Nathan and Laine, Kim and Lauter, Kristin and Naehrig, Michael and Wernsing, John},
  booktitle = 	 {Proceedings of The 33rd International Conference on Machine Learning},
  pages = 	 {201--210},
  year = 	 {2016},
  volume = 	 {48},
  series = 	 {Proceedings of Machine Learning Research},
  address = 	 {New York, New York, USA},
  month = 	 {20--22 Jun},
  publisher =    {PMLR},
  url = 	 {https://proceedings.mlr.press/v48/gilad-bachrach16.html},
}

@misc{100x,
      author = {Wonkyung Jung and Sangpyo Kim and Jung Ho Ahn and Jung Hee Cheon and Younho Lee},
      title = {Over 100x Faster Bootstrapping in Fully Homomorphic Encryption through Memory-centric Optimization with {GPUs}},
      howpublished = {Cryptology ePrint Archive, Paper 2021/508},
      year = {2021},
      url = {https://eprint.iacr.org/2021/508}
}

@misc{RNS,
      author = {Jean-Claude Bajard and Julien Eynard and Anwar Hasan and Vincent Zucca},
      title = {A Full {RNS} Variant of {FV} like Somewhat Homomorphic Encryption Schemes},
      howpublished = {Cryptology ePrint Archive, Paper 2016/510},
      year = {2016},
      url = {https://eprint.iacr.org/2016/510}
}

@misc{Helib,
      author = {Shai Halevi and Victor Shoup},
      title = {Design and implementation of {HElib}: a homomorphic encryption library},
      howpublished = {Cryptology ePrint Archive, Paper 2020/1481},
      year = {2020},
      url = {https://eprint.iacr.org/2020/1481}
}

@misc{Bert,
      title={{BERT}: Pre-training of Deep Bidirectional Transformers for Language Understanding}, 
      author={Jacob Devlin and Ming-Wei Chang and Kenton Lee and Kristina Toutanova},
      year={2019},
      eprint={1810.04805},
      archivePrefix={arXiv},
}

@misc{Lattigo,
    organization = {tuneinsight},
    title = {Lattigo v5},
    howpublished = {Online: \url{https://github.com/tuneinsight/lattigo}},
    month = Nov,
    year = 2023,
    note = {EPFL-LDS, Tune Insight SA}
}

@inproceedings {DaCapo,
author = {Seonyoung Cheon and Yongwoo Lee and Dongkwan Kim and Ju Min Lee and Sunchul Jung and Taekyung Kim and Dongyoon Lee and Hanjun Kim},
title = {{DaCapo}: Automatic Bootstrapping Management for Efficient Fully Homomorphic Encryption},
booktitle = {33rd USENIX Security Symposium (USENIX Security 24)},
year = {2024},
isbn = {978-1-939133-44-1},
address = {Philadelphia, PA},
pages = {6993--7010},
url = {https://www.usenix.org/conference/usenixsecurity24/presentation/cheon},
publisher = {USENIX Association},
month = aug
}

@inproceedings{Cinnamon, author = {Jayashankar, Siddharth and Chen, Edward and Tang, Tom and Zheng, Wenting and Skarlatos, Dimitrios}, title = {Cinnamon: A Framework for Scale-Out Encrypted {AI}}, year = {2025}, isbn = {9798400706981}, publisher = {Association for Computing Machinery}, address = {New York, NY, USA}, url = {https://doi.org/10.1145/3669940.3707260}, doi = {10.1145/3669940.3707260},
booktitle = {Proceedings of the 30th ACM International Conference on Architectural Support for Programming Languages and Operating Systems, Volume 1}, pages = {133–150}, numpages = {18}, location = {Rotterdam, Netherlands}, series = {ASPLOS '25} }

@INPROCEEDINGS{WarpDrive,
  author={Fan, Guang and Zhang, Mingzhe and Zheng, Fangyu and Fan, Shengyu and Zhou, Tian and Deng, Xianglong and Tang, Wenxu and Kong, Liang and Song, Yixuan and Yan, Shoumeng},
  booktitle={2025 IEEE International Symposium on High Performance Computer Architecture (HPCA)}, 
  title={WarpDrive: {GPU}-Based Fully Homomorphic Encryption Acceleration Leveraging Tensor and {CUDA} Cores}, 
  year={2025},
  volume={},
  number={},
  pages={1187-1200},
  doi={10.1109/HPCA61900.2025.00091}}

@INPROCEEDINGS {TensorFHE,
author = { Fan, Shengyu and Wang, Zhiwei and Xu, Weizhi and Hou, Rui and Meng, Dan and Zhang, Mingzhe },
booktitle = { 2023 IEEE International Symposium on High-Performance Computer Architecture (HPCA) },
title = {{TensorFHE: Achieving Practical Computation on Encrypted Data Using GPGPU}},
year = {2023},
volume = {},
ISSN = {},
pages = {922-934},
doi = {10.1109/HPCA56546.2023.10071017},
url = {https://doi.ieeecomputersociety.org/10.1109/HPCA56546.2023.10071017},
publisher = {IEEE Computer Society},
address = {Los Alamitos, CA, USA},
month =mar}

@Misc{Liberate_FHE,
  title={{Liberate.FHE: A New FHE Library for Bridging the Gap between Theory and Practice with a Focus on Performance and Accuracy}},
  author={DESILO},
  year={2023},
  note={\url{https://github.com/Desilo/liberate-fhe}},
}

@Misc{cheddar,
  title        = {Cheddar: A Swift Fully Homomorphic Encryption Library Designed for {GPU} Architectures},
  author       = {Wonseok Choi and Jongmin Kim and Jung Ho Ahn},
  year         = {2024},
  eprint       = {2407.13055},
  archivePrefix= {arXiv},
  primaryClass = {cs.CR},
  doi          = {10.48550/arXiv.2407.13055},
  note         = {15 pages, 8 figures; accepted at ASPLOS 2026}
}

@misc{GLUE,
  title={{GLUE}: A Multi-Task Benchmark and Analysis Platform for Natural Language Understanding},
  author={Wang, Alex and Singh, Amanpreet and Michael, Julian and Hill, Felix and Levy, Omer and Bowman, Samuel R.},
  note={In the Proceedings of ICLR.},
  year={2019}
}

@misc{RTE,
  title={The Fifth {PASCAL} Recognizing Textual Entailment Challenge},
  author={Bentivogli, Luisa and Dagan, Ido and Dang, Hoa Trang and Giampiccolo, Danilo and Magnini, Bernardo},
  booktitle={TAC},
  year={2009}
}

@misc{Rotom,
      author = {Edward Chen and Fraser Brown and Wenting Zheng},
      title = {Bridging Usability and Performance: A Tensor Compiler for Autovectorizing Homomorphic Encryption},
      howpublished = {Cryptology {ePrint} Archive, Paper 2025/1319},
      year = {2025},
      url = {https://eprint.iacr.org/2025/1319}
}

@inproceedings{Orion, author = {Ebel, Austin and Garimella, Karthik and Reagen, Brandon}, title = {Orion: A Fully Homomorphic Encryption Framework for Deep Learning}, year = {2025}, isbn = {9798400710797}, publisher = {Association for Computing Machinery}, address = {New York, NY, USA}, url = {https://doi.org/10.1145/3676641.3716008}, doi = {10.1145/3676641.3716008},
booktitle = {Proceedings of the 30th ACM International Conference on Architectural Support for Programming Languages and Operating Systems, Volume 2}, pages = {734–749}, numpages = {16}, location = {Rotterdam, Netherlands}, series = {ASPLOS '25} }

@misc{Cross,
      title={{Leveraging ASIC AI Chips for Homomorphic Encryption}}, 
      author={Jianming Tong and Tianhao Huang and Leo de Castro and Anirudh Itagi and Jingtian Dang and Anupam Golder and Asra Ali and Jevin Jiang and Arvind and G. Edward Suh and Tushar Krishna},
      year={2025},
      eprint={2501.07047},
      archivePrefix={arXiv},
      primaryClass={cs.CR},
      url={https://arxiv.org/abs/2501.07047}, 
}

@misc{Phantom_FHE,
      author = {Hao Yang and Shiyu Shen and Wangchen Dai and Lu Zhou and Zhe Liu and Yunlei Zhao},
      title = {Phantom: A {CUDA}-Accelerated Word-Wise Homomorphic Encryption Library},
      howpublished = {Cryptology ePrint Archive, Paper 2023/049},
      year = {2023},
      doi = {10.1109/TDSC.2024.3363900},
      note = {\url{https://eprint.iacr.org/2023/049}},
      url = {https://eprint.iacr.org/2023/049}
}

@article{FHEMemory,
   title={Enabling Homomorphically Encrypted Inference for Large {DNN} Models},
   volume={71},
   ISSN={2326-3814},
   url={http://dx.doi.org/10.1109/TC.2021.3076123},
   DOI={10.1109/tc.2021.3076123},
   number={5},
   journal={IEEE Transactions on Computers},
   publisher={Institute of Electrical and Electronics Engineers (IEEE)},
   author={Lloret-Talavera, Guillermo and Jorda, Marc and Servat, Harald and Boemer, Fabian and Chauhan, Chetan and Tomishima, Shigeki and Shah, Nilesh N. and Pena, Antonio J.},
   year={2022},
   month=may, pages={1145–1155} }

@misc{Thor,
      author = {Jungho Moon and Dongwoo Yoo and Xiaoqian Jiang and Miran Kim},
      title = {{THOR}: Secure Transformer Inference with Homomorphic Encryption},
      howpublished = {Cryptology {ePrint} Archive, Paper 2024/1881},
      year = {2024},
      doi = {10.1145/3719027.3765150},
      url = {https://eprint.iacr.org/2024/1881}
}

@misc{FHEFriendlyLLM,
      title={Encryption-Friendly {LLM} Architecture}, 
      author={Donghwan Rho and Taeseong Kim and Minje Park and Jung Woo Kim and Hyunsik Chae and Ernest K. Ryu and Jung Hee Cheon},
      year={2025},
      eprint={2410.02486},
      archivePrefix={arXiv},
      primaryClass={cs.CR},
      url={https://arxiv.org/abs/2410.02486}, 
}

@misc{Bolt,
      author = {Qi Pang and Jinhao Zhu and Helen Möllering and Wenting Zheng and Thomas Schneider},
      title = {{BOLT}: Privacy-Preserving, Accurate and Efficient Inference for Transformers},
      howpublished = {Cryptology {ePrint} Archive, Paper 2023/1893},
      year = {2023},
      url = {https://eprint.iacr.org/2023/1893}
}

@inproceedings{Neo, author = {Jiao, Dian and Deng, Xianglong and Wang, Zhiwei and Fan, Shengyu and Chen, Yi and Meng, Dan and Hou, Rui and Zhang, Mingzhe}, title = {Neo: Towards Efficient Fully Homomorphic Encryption Acceleration using Tensor Core}, year = {2025}, isbn = {9798400712616}, publisher = {Association for Computing Machinery}, address = {New York, NY, USA}, url = {https://doi.org/10.1145/3695053.3731408}, doi = {10.1145/3695053.3731408}, 
booktitle = {Proceedings of the 52nd Annual International Symposium on Computer Architecture}, pages = {107–121}, numpages = {15}, location = { }, series = {ISCA '25} }

@inproceedings{Mad, author = {Agrawal, Rashmi and De Castro, Leo and Juvekar, Chiraag and Chandrakasan, Anantha and Vaikuntanathan, Vinod and Joshi, Ajay}, title = {{MAD}: Memory-Aware Design Techniques for Accelerating Fully Homomorphic Encryption}, year = {2023}, isbn = {9798400703294}, publisher = {Association for Computing Machinery}, address = {New York, NY, USA}, url = {https://doi.org/10.1145/3613424.3614302}, doi = {10.1145/3613424.3614302}, 
booktitle = {Proceedings of the 56th Annual IEEE/ACM International Symposium on Microarchitecture}, 
pages = {685–697}, numpages = {13}, location = {Toronto, ON, Canada}, series = {MICRO '23} }

@misc{tsmc_cowos,
  author = {{Taiwan Semiconductor Manufacturing Company Limited}},
  title = {CoWoS®: Chip on Wafer on Substrate Technology},
  howpublished = {\url{https://3dfabric.tsmc.com/english/dedicatedFoundry/technology/cowos.htm}},
  year = {2024},
}

@article{gouert2025hardware,
  title={Hardware-accelerated encrypted execution of general-purpose applications},
  author={Gouert, Charles and Joseph, Vinu and Dalton, Steven and Augonnet, Cedric and Garland, Michael and Tsoutsos, Nektarios Georgios},
  journal={Proceedings on Privacy Enhancing Technologies},
  year={2025}
}

@inproceedings{GME,
  title={{GME}: {GPU}-based microarchitectural extensions to accelerate homomorphic encryption},
  author={Shivdikar, Kaustubh and Bao, Yuhui and Agrawal, Rashmi and Shen, Michael and Jonatan, Gilbert and Mora, Evelio and Ingare, Alexander and Livesay, Neal and Abell{\'a}n, Jos{\'e} L and Kim, John and others},
  booktitle={Proceedings of the International Symposium on Microarchitecture},
  pages={670--684},
  year={2023}
}

@inproceedings{doubleHoisting, author = {Bossuat, Jean-Philippe and Mouchet, Christian and Troncoso-Pastoriza, Juan and Hubaux, Jean-Pierre}, title = {Efficient Bootstrapping for Approximate Homomorphic Encryption with Non-sparse Keys}, year = {2021}, isbn = {978-3-030-77869-9}, publisher = {Springer-Verlag}, address = {Berlin, Heidelberg}, url = {https://doi.org/10.1007/978-3-030-77870-5_21}, doi = {10.1007/978-3-030-77870-5_21},
booktitle = {Advances in Cryptology – EUROCRYPT 2021: 40th Annual International Conference on the Theory and Applications of Cryptographic Techniques, Zagreb, Croatia, October 17–21, 2021, Proceedings, Part I}, pages = {587–617}, numpages = {31}, location = {Zagreb, Croatia} }

@misc {Mage,
author = {Sam Kumar and David E. Culler and Raluca Ada Popa},
title = {{MAGE}: Nearly Zero-Cost Virtual Memory for Secure Computation},
booktitle = {15th {USENIX} Symposium on Operating Systems Design and Implementation ({OSDI} 21)},
year = {2021},
isbn = {978-1-939133-22-9},
pages = {367--385},
url = {https://www.usenix.org/conference/osdi21/presentation/kumar},
publisher = {{USENIX} Association},
month = jul
}

@misc{FIDESlib,
  title={{FIDESlib: A Fully-Fledged Open-Source FHE Library for Efficient CKKS on GPUs}},
  author={Agull{\'o}-Domingo, Carlos and Vera-L{\'o}pez, {\'O}scar and Guzelhan, Seyda and Daksha, Lohit and El Jerari, Aymane and Shivdikar, Kaustubh and Agrawal, Rashmi and Kaeli, David and Joshi, Ajay and Abell{\'a}n, Jos{\'e} L},
  booktitle={Proceedings of the International Symposium on Performance Analysis of Systems and Software (ISPASS)},
  pages={1--3},
  year={2025},
}

@misc{NCCL,
  title = {{NVIDIA} Collective Communications Library ({NCCL})},
  author = {{NVIDIA}},
  howpublished = {\url{https://developer.nvidia.com/nccl}},
  year = {2025}
}

@misc{bfv1,
      author = {Junfeng Fan and Frederik Vercauteren},
      title = {Somewhat Practical Fully Homomorphic Encryption},
      howpublished = {Cryptology {ePrint} Archive, Paper 2012/144},
      year = {2012},
      url = {https://eprint.iacr.org/2012/144}
}

@misc{bfv2, author = {Brakerski, Zvika}, title = {Fully Homomorphic Encryption without Modulus Switching from Classical GapSVP}, year = {2012}, isbn = {9783642320088}, publisher = {Springer-Verlag}, address = {Berlin, Heidelberg}, url = {https://doi.org/10.1007/978-3-642-32009-5_50}, doi = {10.1007/978-3-642-32009-5_50},pages = {868–886}, numpages = {19} }

@article{heprs,
  author = {Knight, Elizabeth and Li, Jiaqi and Jensen, Matthew and Yolou, Israel
            and Kockan, Can and Gerstein, Mark},
  title = {Homomorphic Encryption Enables Privacy Preserving Polygenic Risk Scores},
  journal = {Cell Reports Methods},
  volume = {6},
  year = {2026},
  doi = {10.1016/j.crmeth.2025.101271}
}

@article{hefa,
  author = {Froelicher, David and Troncoso-Pastoriza, Juan R. and Raisaro,
            Jean Louis and Cuendet, Michel A. and Sa Sousa, Joao and Cho,
            Hyunghoon and Berger, Bonnie and Fellay, Jacques and Hubaux,
            Jean-Pierre},
  title = {Truly Privacy-Preserving Federated Analytics for Precision Medicine
           with Multiparty Homomorphic Encryption},
  journal = {Nature Communications},
  volume = {12},
  year = {2021},
  doi = {https://doi.org/10.1038/s41467-021-25972-y}
}

@article{akavia2019search,
  author = {Akavia, Adi and Gentry, Craig and Halevi, Shai and Leibovich, Max},
  title = {Setup-Free Secure Search on Encrypted Data: Faster and
           Post-Processing Free},
  journal = {Proceedings on Privacy Enhancing Technologies},
  volume = {2019},
  number = {3},
  pages = {87--107},
  year = {2019},
  doi = {10.2478/popets-2019-0038}
}

@inproceedings{kim2026ppvul,
  author = {Kim, Seungho and Park, Seonhye and Kim, Jihun and Kim, Eunsoo
            and Oh, Sanghak and Choi, Hyunmin and Kim, Hyoungshick},
  title = {{PP-Vul}: Privacy-Preserving Vulnerability Detection Using
           Homomorphic Encryption},
  booktitle = {Proceedings of the ACM Asia Conference on Computer and
               Communications Security},
  pages = {1415--1430},
  year = {2026},
  doi = {10.1145/3779208.3785415}
}

@techreport{oecd2025pets,
  author = {{OECD}},
  title = {Sharing Trustworthy {AI} Models with Privacy-Enhancing Technologies},
  institution = {OECD Publishing},
  series = {OECD Artificial Intelligence Papers},
  number = {38},
  year = {2025},
  doi = {10.1787/a266160b-en}
}

@misc{nistFHE,
  author = {{National Institute of Standards and Technology}},
  title = {Privacy-Enhancing Cryptography: Fully-Homomorphic Encryption ({FHE})},
  year = {2026},
  url = {https://csrc.nist.gov/projects/pec/fhe},
  note = {Accessed: 2026-08-26}
}

@article{
heOncology,
author = {Ravit Geva  and Alexander Gusev  and Yuriy Polyakov  and Lior Liram  and Oded Rosolio  and Andreea Alexandru  and Nicholas Genise  and Marcelo Blatt  and Zohar Duchin  and Barliz Waissengrin  and Dan Mirelman  and Felix Bukstein  and Deborah T. Blumenthal  and Ido Wolf  and Sharon Pelles-Avraham  and Tali Schaffer  and Lee A. Lavi  and Daniele Micciancio  and Vinod Vaikuntanathan  and Ahmad Al Badawi  and Shafi Goldwasser },
title = {Collaborative privacy-preserving analysis of oncological data using multiparty homomorphic encryption},
journal = {Proceedings of the National Academy of Sciences},
volume = {120},
number = {33},
pages = {e2304415120},
year = {2023},
doi = {10.1073/pnas.2304415120},
URL = {https://www.pnas.org/doi/abs/10.1073/pnas.2304415120},
}
